\documentclass[journal]{IEEEtran}
\usepackage{graphicx}
\usepackage{graphicx}
\usepackage{epstopdf}
\usepackage{cite}
\usepackage{amsfonts}
\usepackage{amssymb}
\usepackage{amsmath}
\usepackage{mathrsfs}
\usepackage{bm}
\usepackage{amsmath}
\usepackage{mdwmath}
\usepackage{mdwtab}
\usepackage{subfigure}
\usepackage{dsfont}
\usepackage[ruled,vlined,linesnumbered]{algorithm2e} 

\DeclareMathOperator*{\argmax}{arg\,max}
\DeclareMathOperator*{\argmin}{arg\,min}

\newtheorem{remark}{Remark}
\newtheorem{theorem}{Theorem}
\newtheorem{coro}{Corollary}
\newtheorem{lemma}{Lemma}

\newtheorem{definition}{Definition}

\begin{document}

	\title{Task Replication for Deadline-Constrained Vehicular Cloud Computing: Optimal Policy, Performance Analysis and Implications on Road Traffic}
	\author{Zhiyuan Jiang,~\IEEEmembership{Member,~IEEE}, Sheng Zhou,~\IEEEmembership{Member,~IEEE}, Xueying Guo,~\IEEEmembership{Member,~IEEE},\\ Zhisheng Niu,~\IEEEmembership{Fellow,~IEEE}
    \thanks{
    Z. Jiang, S. Zhou and Z. Niu are with Tsinghua National Laboratory for Information Science and Technology, Tsinghua University, Beijing 100084, China. Emails: \{zhiyuan, sheng.zhou, niuzhs\}@tsinghua.edu.cn. 
    X. Guo is with the Department of Computer Science, University of California at Davis, Davis, CA 95616, USA. Email: guoxueying@outlook.com. The corresponding author is S. Zhou. 
    
    This work is sponsored in part by the Nature Science Foundation of China (No. 61701275, No. 91638204, No. 61571265, No. 61621091), the China Postdoctoral Science Foundation, and Intel Collaborative Research Institute for Mobile Networking and Computing. 
    }}
	\maketitle
	
	\begin{abstract}
		In vehicular cloud computing (VCC) systems, the computational resources of moving vehicles are exploited and managed by infrastructures, e.g., roadside units, to provide computational services. The offloading of computational tasks and collection of results rely on successful transmissions between vehicles and infrastructures during encounters. In this paper, we investigate how to provide timely computational services in VCC systems. In particular, we seek to minimize the deadline violation probability given a set of tasks to be executed in vehicular clouds. Due to the uncertainty of vehicle movements, the task replication methodology is leveraged which allows one task to be executed by several vehicles, and thus trading computational resources for delay reduction. The optimal task replication policy is of key interest. We first formulate the problem as a finite-horizon sampled-time Markov decision problem and obtain the optimal policy by value iterations. To conquer the complexity issue, we propose the balanced-task-assignment (BETA) policy which is proved optimal and has a clear structure: it always assigns the task with the minimum number of replicas. Moreover, a tight closed-form performance upper bound for the BETA policy is derived, which indicates that the deadline violation probability follows the Rayleigh distribution approximately. Applying the vehicle speed-density relationship in the traffic flow theory, we find that vehicle mobility benefits VCC systems more compared with road traffic systems, by showing that the optimum vehicle speed to minimize the deadline violation probability is larger than the critical vehicle speed in traffic theory which maximizes traffic flow efficiency. 
	\end{abstract}
	\begin{IEEEkeywords}
    Vehicular networks, cloud computing, Markov decision process, task replication, hard deadline.
    \end{IEEEkeywords}
	
	\section{Introduction}
	\label{sec_intro}
	With the rapid development of autonomous driving technologies and a wide variety of on-board infotainment services, the future vehicles are equipped with advanced and sufficient computational resources to facilitate sophisticated artificial-intelligence (AI) based algorithms. Wireless communication modules are also indispensable, which enable various types of communications including vehicle-to-vehicle (V2V), vehicle-to-infrastructure (V2I) and vehicle-to-everything (V2X) over different protocols, e.g., dedicated short range communication (DSRC) \cite{kenney11}, long-term-evolution-vehicle (LTE-V) \cite{chen16} and future $5$G technologies \cite{wang14}. Armed with these components, the vehicular network is evolving into a connected group of intelligent individuals which has great potentials. Consequently, the vehicle-as-a-resource (VaaR) concept has been proposed recently which exploits the enormous resources of vehicles and the vehicular network to facilitate new types of services such as sensing, vehicle-bearing data transfer, cloud computing and localization \cite{hou16,lin17,zheng15, Jang17,bitam15,abd15,gerla12,lee14,whai14,hus12,step11}. 
	\begin{figure*}[!t]
		\centering
		\includegraphics[width=0.85\textwidth]{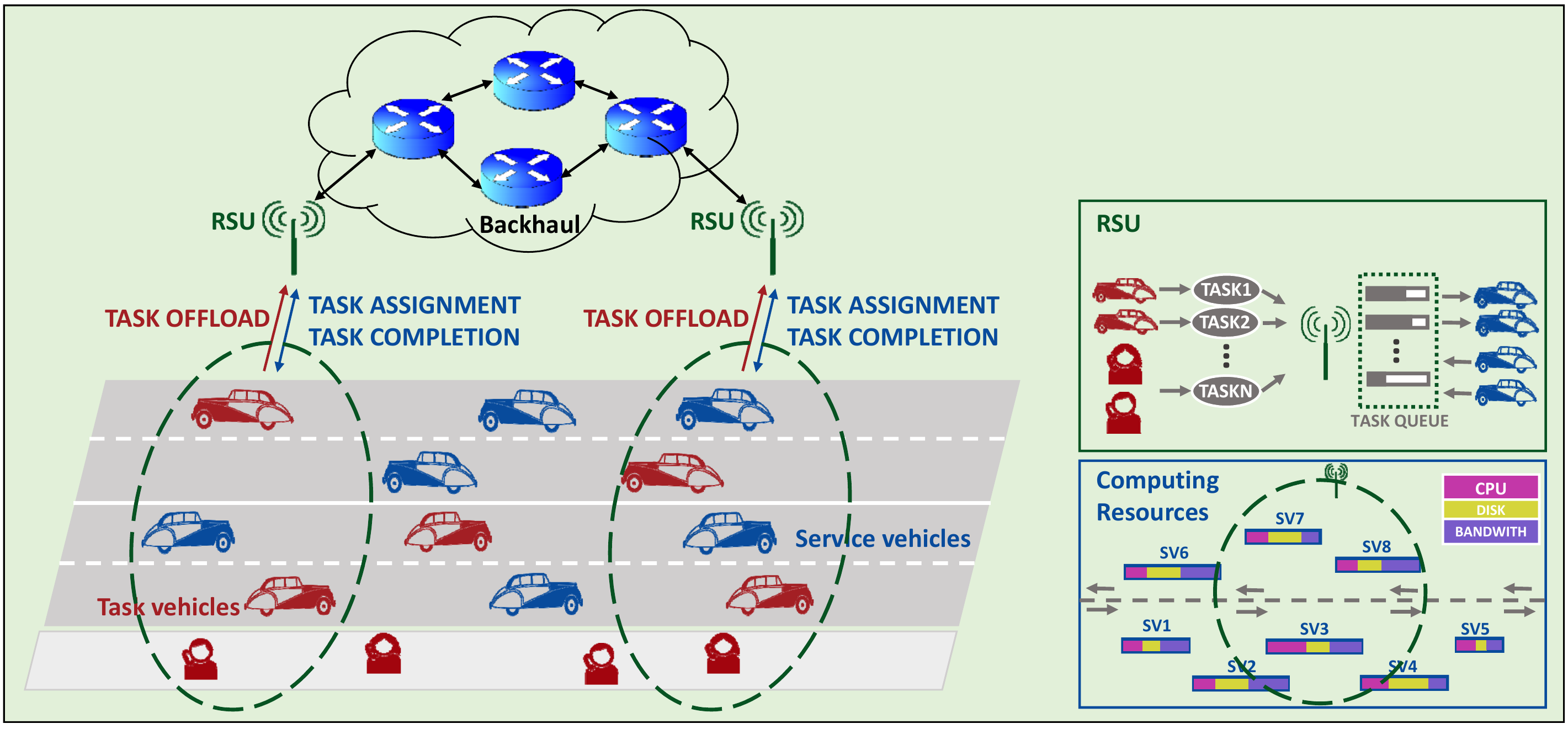}
		\caption{Vehicular cloud computing system overview.}
		\label{Fig_sys}
	\end{figure*}
	
	Among the resources that the vehicular network can offer, the computational resources are considered to be the most desirable one due to the growing computing demand of AI systems and the hardware limitation of user equipments \cite{Jang17} \cite{bitam15}. The vehicular cloud computing (VCC) \cite{hou16} \cite{abd15} is therefore proposed to allow the computing tasks to be collected at the task assignment station, e.g., a roadside unit (RSU), and offloaded to vehicles. A pictorial description of the VCC architecture is given in Fig. \ref{Fig_sys}, in which the RSU controls the vehicular cloud through communications with vehicles inside its coverage area \cite{mer13}. The computational resources of vehicles are abstracted based on the network function virtualization (NFV) technology \cite{han15nfv}, such that it is transparent to users. The users can be pedestrians, sensors, and vehicles with insufficient computation capability. The task execution procedure is as follows. An RSU, denoted by task-RSU, collects tasks from users and offloads them to its associated vehicles based on a given task assignment policy during meetings between vehicles and the RSU. After the task computation, the output of the task is fed back to the task-RSU through the backhaul when the vehicle meets with any RSU in the network. Since the tasks can only be offloaded (or collected) to (or from) the vehicles inside the limited coverage area by successful transmissions during encounters \cite{mer13}, such a VCC system is referred to as an encounter-based VCC (eVCC) system \cite{yousef08,zhou11,zhou13,zhou16}. One can immediately identify that the most intriguing challenge in the eVCC system is that the service process exhibits large uncertainty due to vehicle mobility. The meetings between the task-RSU and service-vehicles (SVs) and the meetings afterwards when the outputs of the tasks are collected are both subject to the random behaviour of moving vehicles. Consequently, the timeliness of the task computation is difficult to guarantee. Therefore, how to provide timely cloud computing services in eVCC systems becomes a challenging problem. 
	
	We propose to exploit the task replication technique to overcome this problem, which allows one task to be offloaded to multiple vehicles and the task is executed as long as one of the vehicles executes the task and feeds back the output. In this way, the mobility and high volume of vehicles are, to some extent, exploited to provide low-latency services. Towards this end, the key questions include how to assign (replicate) tasks to minimize the deadline violation probability of tasks given stochastic vehicle movements, and how to analyze and optimize the performance.
	
	In the literature, there are some recent efforts dedicated to VCC systems. In \cite{eck11} \cite{liu11}, the authors propose to exploit parked cars for task offloading and cooperative sensing. On the contrary, since parked cars can be regarded as part of the static cloud servers and the corresponding task offloading policy has been well investigated (see \cite{kumar10} \cite{you17} and references therein), we focus on utilizing vehicles on the move. The eVCC task scheduling and resource allocation problem are formulated based on a semi-Markov-decision-process (SMDP) approach in \cite{zheng15} \cite{liang12}. However, the SMDP based approach suffers from the curse of dimensionality and hence cannot be applied in high vehicle density scenarios. A large body of work has focused on the task replication policy and performance analysis in, e.g., data retrieval, multi-server data processing systems \cite{gard15,shah16,sun15,wang15,kim09,dean13}. The task replication technique is usually leveraged to deal with the straggler problem \cite{wang15} \cite{dean13}, where the service process has a heavy-tail distribution. In \cite{gard15}, the authors focus on analyzing the performance of task replication policies based on the queuing theory. In \cite{shah16} \cite{kim09}, the optimal replication degree, i.e., the number of replicas, is investigated. However, it is not applicable to the eVCC system where the servers (vehicles) are not always present. In \cite{sun15}, the optimal data retrieval replication policy is proved to be an shortest expected remaining processing time policy, which essentially schedules the task with the shortest time to go. Interestingly, although in different contexts, we will show that the optimal policy to minimize the deadline violation probability in VCC systems actually schedules the task with the longest time to go (the task with least number of replicas has the longest expected time to go).
	
	In this paper, the optimal multi-task replication policy to minimize the deadline violation probability and its performance in the eVCC system are investigated. The main contributions include:
	\begin{itemize}
	    \item 
	    We first formulate the task assignment problem as a sampled-time MDP, by which the optimal policy and deadline violation probability can be computed by finite-horizon value iteration. This result is used as the benchmark to validate optimality. To conquer the computational complexity issue, we design the balanced-task-assignment (BETA) policy by which the resultant task assignment is as balanced as possible, and prove its optimality.
	    \item
	    A tight upper bound for the deadline violation probability following the optimal BETA policy is derived in closed-form, which implies that the optimum deadline violation probability follows the Rayleigh distribution approximately. We further apply the results from the vehicular traffic flow theory and demonstrate that there exists an interesting relationship between computation delay and vehicle density (or speed). A larger vehicle density leads to more vehicles executing the tasks, but meanwhile slows down the average vehicle speed due to limited human reaction time \cite{rose11} which results in larger computation delay considering the task output has to be collected after offloaded. We show that the optimum vehicle density to minimize the deadline violation probability is in general lower, and the corresponding vehicle speed is higher, than the critical vehicle density and speed in traffic theory to maximize traffic efficiency, respectively.
	\end{itemize}
	
	The remainder of the paper is organized as follow. In Section \ref{sec_sm}, the system model is presented and the problem is formulated. In Section \ref{sec_mdp}, a sampled-time MDP based solution is presented and the results are utilized as a benchmark to validate policy optimality. In Section \ref{sec_bta}, the optimal BETA policy is described and proved for optimality. In Section \ref{sec_pa}, a closed-form performance bound is derived for the deadline violation probability given the optimal BETA policy and the optimum vehicle density is analyzed by applying the results from the vehicular traffic theory in Section \ref{sec_imp}. The simulation results are conveyed in Section \ref{sec_nr}. Finally, we conclude our work in Section \ref{sec_cl}.
	
	Throughout the paper, we use boldface uppercase letters, boldface lowercase letters and lowercase letters to designate matrices, column vectors and scalars, respectively. Denote by $\mathbb{E}(\cdot)$ as the expectation operation. Denote by $\mathds{1}(\cdot)$ as the indicator function. The exponential distribution with mean of $\lambda$ is denoted as $\textrm{Exp}(\lambda)$, and $\exp(\cdot)$ denotes the exponential function. The cardinality of a set is denoted by $|\cdot|$. The probability of an event is denoted by $\Pr\{\cdot\}$. Denote by $k\,\textrm{mod}\,N$ as the remainder of the Euclidean division of $k$ by $N$.
	
	\section{System Model and Problem Formulations}
	\label{sec_sm}
	We consider a Poisson vehicular network model with exponential vehicle-meeting time (PV-E). The following two model assumptions are made.
	\begin{enumerate}
		\item 
		\emph{Poisson vehicle spatial distribution}: The number of vehicles in the system at time $t$, denoted by $M_t$, follows the Poisson distribution\footnote{The Poisson point process is widely-used to model the stochastic spatial geometry of, e.g., base stations and vehicles \cite{chiu13}.}. Denoting the length of roads in the system as $S$, the average vehicle density is $\lambda$, then
		\begin{equation}
		\Pr\left\{M_t=k\right\} = \frac{\lambda^k S^k e^{-\lambda S}}{k!}.
		\end{equation}
		The vehicle number is assumed to be the same during the task offloading since the deadline is usually much smaller than the vehicle number variation time in vehicular networks. 
		\item
		\emph{Homogeneous memory-less vehicle meeting}: At any time and any RSU in the system, the time remaining before vehicle-$m$ meets\footnote{In this paper, we define a meeting between a vehicle and an RSU as an event that a transmission (task offloading or output collection) is successful between them during the passage of the vehicle through the coverage area of the RSU. This definition excludes the encounters during which the transmissions are unsuccessful due to, e.g., bad wireless channel conditions, insufficient passage time and etc. } with the RSU follows an independent exponential distribution with mean $\mu_m=\mu$, $\forall m$,\footnote{The exponential meeting time of vehicles is validated and widely-used to model the vehicle behavior \cite{yousef08,Chandra08,zhou11, allen14}.} i.e.,
		\begin{equation}
		\label{homo_meet}
    	\Pr\{t<x\} = 1-e^{- \mu x}.
    	\end{equation}
	\end{enumerate}

	Consider a task-RSU with $N$ tasks that need to be offloaded and executed before deadline $D_n$, $n\in\{1,...,N\}$ where $n$ is the task index. The task size is normalized such that each task can be offloaded to one vehicle at a time. In this paper, we consider the tasks are the with same type, i.e., the delay deadline is the same for all tasks and hence hereinafter $D_n=D$.\footnote{The following MDP formulation can be immediately generalized to the multi-type task scenario, but the analysis for the optimal policy would be considerably more complicated and is left for future work.} The tasks are executed in the eVCC system as follows. For each task, it has to be first offloaded to a vehicle when the vehicle meets with the task-RSU. After the meeting, the task is assumed to be successfully offloaded and the vehicle uses on-board computational resources to compute the task. After that, when the vehicle meets with any RSU in the system, the computing result is fed back to the task-RSU through backhaul and the task is considered as successfully executed. Moreover, we denote 
	\begin{equation}
	{\omega _n} = \left\{ {\begin{array}{*{20}{lr}}
      { 1}, \textrm{ if the task-$n$ is executed before deadline} \\ 
      { 0}, \textrm{ otherwise}
    \end{array}} \right.
	\end{equation}
	In this paper, we focus on the scenario where the task computation time is negligible compared with vehicle meeting time as shown in \cite{zheng15}. 
	
	We consider the task assignment problem that upon the meeting between a vehicle and the task-RSU, which task should be offloaded to the vehicle so as to minimize the deadline violation ratio, which is defined as
	\begin{equation}
	\label{pv}
	p_\textrm{v} = 1-\sum_{n=1}^N \frac{\omega_n}{N}.
	\end{equation}
	
	Moreover, in the eVCC system, the task service time fluctuates dramatically due to the unpredictable and diversified vehicle movements. Therefore, in order to counteract this so-called straggler effect \cite{dean13}, the tasks are replicated and executed independently. One task can be offloaded multiple times to different vehicles. Once one of the vehicles finishes the task (the output is fed back), the task is considered to be successfully executed. Since the SVs are out of coverage area when not meeting with RSUs in the eVCC system, they cannot be assigned new tasks even if the task at hand has been finished by other vehicles, until they meet with an RSU again. This is referred to as the non-purging model in \cite{pol16}. In a purging model for e.g., multi-processor computing, the processor begins a new task as soon as the current task is finished (maybe by other processors).
	
	\subsection{Task Execution Procedure in EVCC Systems}
	The detailed task execution procedure in eVCC systems is described in the following. 

	    \textbf{Task collection:} Define the task execution interval as $T$. Within the interval, the task collection and assignment are executed simultaneously. The task requests from users are collected at each RSU. Denote the number of collected task requests at the $t$-th interval as $N_t$.
	    
	    \textbf{Task Assignment:} In the $(t+1)$-th interval, the $N_t$ tasks collected in the previous interval at each RSU are assigned based on the proposed algorithm described in Section \ref{sec_bta}. Specifically, upon the arrival of a vehicle, the $n$-th task is assigned based on the algorithm. If the assignment is successful subject to wireless transmission failure, the number of replications for the $n$-th task increases by one.\footnote{Note that each RSU carries out the task assignment independently, which means that the recourse constraints of vehicle are ignored in this paper.} 
	    
	    \textbf{Task Output Feedback:} At the end of each interval, each RSU collects the task output feedback from assigned vehicles through backhaul links. The assigned vehicles may finish computing the tasks and feed back the task outputs when meeting with another RSU. In this case the output is transmitted to the original RSU through backhaul links. Afterwards, each RSU transmits the outputs to corresponding users and the task execution is finished. The above procedure then repeats itself. 
	    
	\section{Sampled-Time Markov Chain Based Approach}
	\label{sec_mdp}
	First, the continuous time model is converted to a discrete-time approximation. Consider some given small increment of time $\delta$, a sampled time approximation to the continuous time Poisson vehicle meeting for vehicle-$m$ is a discrete Bernoulli process whose meeting probability is $\mu_m \delta$ \cite{gal12}. It is straightforward to derive that the meeting rate for a total of $K$ vehicles is $\sum_{m=1}^K \mu_m$, and thus the meeting probability is $K \mu \delta$ due to the second assumption in Section \ref{sec_sm}. 
	
	On account of the task replication, the system state space is characterized by 	
	\begin{equation}
	\label{state}
	\mathcal{S} =\left\{\left(r_{1},...,r_{N},d\right)|1 \le d \le D,\,r_i \in \Gamma, \forall i=1,...,N\right\},
	\end{equation}
	where $d$ is the number of elapsed time slots, i.e., the computation delay. Note that the problem is with finite time horizon of $1 \le d \le D$, $d \in \mathbb{Z}$, for deadline violation consideration. The number of current replicas of task-$n$ which are being processed in the eVCC system is denoted by $r_{n}$, which is sufficient to describe the system state due to the memory-less vehicle meeting assumption. The time slot each replica is made is irrelevant based on the assumption. The state space is finite since the number of replicas cannot exceed the number of vehicles in the system. For ease of exposition, with slight abuse of notations we denote $r_{n} = \textrm{F}$ as that task-$n$ is finished. The set 
	\begin{equation}
	\Gamma=\{1,..., M,\textrm{F}\}
	\end{equation}
	denotes the set of values of $r_{n}$, $\forall n$. For each state with $d=D$, it denotes a terminal state. Since $1 \le d \le D$ and the terminal states are mutually exclusive, there exists a unique terminal state where the system terminates in. The reward is defined as the number of finished tasks at time $D$. Based on this definition and \eqref{pv}, maximizing the reward is equivalent to minimizing the deadline violation ratio. Also note that the initial state is always 
	\begin{equation}
	    s_0 = (\underbrace {0,...,0}_N,0).
	\end{equation}
	It is without loss of generality to assume there is a vehicle meeting at the time slot zero, since it is equivalent to consider the delay deadline as $D-D_0$ if the first time when a vehicle meets with the task-RSU is $D_0$.
	
	The system state is reviewed and the decisions are made at the meeting time\footnote{Although the meeting is defined as an event that a successful transmission occurs during the passage of the vehicle through the coverage area of an RSU, the meeting time is defined as a time point instead of a period of time at which the transmission is completed.} between vehicles and the task-RSU, expect the time slot zero. Note that based on the discrete time conversion, a sufficiently small time increment $\delta$ is chosen such that at any given time slot the probability of multiple vehicle meetings is negligible. It is easy to validate the fact that based on the system model and system state description in \eqref{state}, the following Markovian properties are satisfied: If an action $a$ is chosen in state $x$ at a decision time slot, then the time till the next decision time slot, i.e., $\tau$, is determined by the memory-less vehicle meetings with the RSU. Moreover, the system state at the next decision time slot $s_{t+\tau}$ depends on the task execution during time interval $(t,\,t+\tau)$ which is also memory-less. Hence, the time till next decision time slot and the next system state depend only on the present state $x$ and the chosen action $a$, i.e.,
	\begin{IEEEeqnarray}{rCl}
	&& \Pr\{s_{t+\tau} = y|(s_t=x,a_t=a),...,(s_0,a_0)\} \nonumber\\
	&=& \Pr\{s_{t+\tau}=y|(s_t=x,a_t=a)\}, \nonumber\\
	&& \Pr\{\tau=z |(s_t=x,a_t=a),...,(s_0,a_0)\} \nonumber\\
	&=& \Pr\{\tau=z|(s_t=x,a_t=a)\}.
	\end{IEEEeqnarray}
	Also the reward incurred only depends on the terminal state, and thus independent of the past. 
	
	\subsection{Transition Probability}
	At the decision time slot, given an action that the $n$-th task (unfinished) is offloaded to the meeting vehicle, the state transition probability is
	\begin{IEEEeqnarray}{rCl}
		\label{pr}
		&&\Pr\{((r_i|i\in \mathcal{M}, i\neq n),r_n,(r_j|j\notin \mathcal{M}, j\neq n ),d ) \to \nonumber \\ 
		&&((\textrm{F}|i \in \mathcal{M}, i\neq n), r_{n}+1, (r_j| j\notin \mathcal{M}, j\neq n)), d+\tau\} = \nonumber\\
		&& \left\{ {\begin{array}{*{20}{lr}}
				(1-s)^{\tau-1}s \prod_{n\notin \mathcal{M}} p_n^{\tau} \prod_{m\in \mathcal{M}}(1-p_m^{\tau}),& d+\tau <D; \nonumber\\
				(1-s)^{\tau-1} \prod_{n\notin \mathcal{M}} p_n^{\tau} \prod_{m\in \mathcal{M}}(1-p_m^{\tau}),& d+\tau =D, 
		\end{array}} \right.\\
	\end{IEEEeqnarray}
    where $p_i$ denotes the probability that the $i$-th task is not finished until the next decision time slot. The set $\mathcal{M}$ contains the tasks that have not been finished before time $d$ and are finished during time slot $d$ to the next decision time slot. It follows that
	\begin{equation}
	\label{p_i}
	p_i=\max\{1-q_i,\,\mathds{1}(r_i=\textrm{F})\},\,i\in\{1,\cdots,N\}
	\end{equation}
	and	
	\begin{equation}
	\label{q_i}
	q_i=(r_i+\mathds{1}(i=n))B\mu\delta, \, i\in\{1,\cdots,N\},
	\end{equation}
	where $q_i$ denotes the probability that at least one of the vehicles which are executing task-$i$ meets with one of the $B$ RSUs. Note that there are $\left(r_i+\mathds{1}(i=n)\right)B$ different vehicle-RSU combinations that can lead to the execution of the $i$-th task. Based on the homogeneous memory-less vehicle meeting assumption, the event that task-$i$ is finished happens with a rate of $\left(r_i+\mathds{1}(i=n)\right)B\mu$, due to the fact that the distribution of the minimum of several exponentially distributed random variables is also exponential with a rate that is the summation of the rates of the variables. Denote the probability that a vehicle meets with the task-RSU as $s$. Based on the same arguments,
	\begin{equation}
	\label{sss}
	s = M \mu \delta,
	\end{equation}
	where $M$ denotes the number of vehicles in the system during this task computation. Based on the first assumption in Section \ref{sec_sm}, $M$ is Poisson distributed but fixed during one task computation. Therefore, the time index is omitted for simplicity and its stochastic nature will be considered in the performance analysis in Section \ref{sec_pa}. Note that $\delta$ is sufficiently small such that $s$ is also small due to the discrete time approximation formulation. Assuming the current decision time slot is $d$, the probability that the next vehicle meets with the task-RSU at time $d+\tau$ is 
	\begin{equation}
	\Pr\{\textsl{next meeting at time }d+\tau\} = (1-s)^{\tau-1}s,
	\end{equation}
	when $d+\tau<D$. Since a finite horizon of $1 \le d \le D$ is sufficient, the transition probability to a terminal state, which is denoted by the time index of $D$, is 
	\begin{equation}
	\Pr\{\textsl{no more meeting before deadline}\} = (1-s)^{D-d-1}.
	\end{equation}
	We assume that the vehicle meetings are independent. Therefore, the task completion and the new vehicle meetings are independent. 
	
	\subsection{Finite-Horizon MDP-Based Solution}
	
	The system reward is expressed by a terminal reward vector $\bm{h}$, where the entry of $\bm{h}$ is a function of a terminal state $s_\textrm{t} \in \mathcal{S}$, i.e.,
	\begin{equation}
	\label{h(s)}
	h(s_\textrm{t})=\sum_{i=1}^{N}\mathds{1}(s_\textrm{t}(i)=\textrm{F}),
	\end{equation}
	which denotes the number of finished tasks before deadline. Consequently, the solution to the formulated MDP minimizes the deadline violation ratio in \eqref{pv}.
	
	Since this is a finite horizon MDP problem with finite state space, the backwards recursive iteration algorithm can be applied to find the optimal dynamic policy. Note that the optimal policy of a finite horizon problem is, in general, a dynamic policy, i.e., not a stationary policy as often observed in the infinite-horizon MDP, meaning that given the current system state, the policy also depends on the current time. Applying the backwards recursive iteration, we start from the terminal stage, denoted by stage $K$
	\begin{equation}
	\label{J_k}
	J_k(s_k)=\max_{a_k \in \{1,...,N\}}\mathbb{E}\{J_{k+1}(s_{k+1})+g_k(s_k)\},
	\end{equation}
	where $J_K(s_K)=g_K(s_K)$ is the terminal reward. Based on \cite{bersk}, the optimal reward 
	\begin{equation}
	J^{\ast}(s_0)=J_0(s_0),
	\end{equation}
	and the optimal dynamic policy is given by $a^{\ast}=\{a_0^{\ast},\cdots,a_{K-1}^{\ast}\}$, where $a_k^{\ast}$'s are derived from \eqref{J_k}.
	
	\section{Optimal Balanced-Task-Assignment Policy}
	\label{sec_bta}	
	The finite-horizon MDP-based solution described above is guaranteed to find the optimum. However, as in many application scenarios, it suffers from the curse of dimensionality. One can easily find that the state space grows exponentially with the number of tasks. In what follows, we prove that the BETA policy, which is also stationary, is the optimal policy. First, the definition of the BETA policy is given. 
	
	\begin{definition}
		\label{def-1}
		The BETA policy is defined as at every decision time slot when a vehicle meets with the task-RSU, given the current task status $(r_1,\cdots, r_N)$, offload the $n$-th task to the vehicle, where
		\begin{equation}
		\label{offload-n}
		n=\argmin_{i \in \{1,...,N\}}\{r_i | r_i\neq \textrm{F}\},
		\end{equation}
		and ties are broken arbitrarily. 
	\end{definition}	
	\begin{remark}
		It is not difficult to figure out that the task status following the BETA policy is as balanced as possible, in the sense that given the total number of computing vehicles as $\gamma=\sum_{r_i \neq \rm{F}}r_i$, and the number of unfinished tasks as $u=|\{r_i| r_i\neq \textrm{F}\}|$, 
		\begin{equation}
		r_i^{\ast} \in \{ \lfloor \gamma/u\rfloor,\lceil \gamma/u\rceil\},\,i\in\{1,\cdots,N\},
		\end{equation}
		where $r_i^{\ast}$ denotes the resultant task assignment by the BETA policy, and $\lfloor x \rfloor$ and $\lceil x \rceil$ denote the nearest integers below and above $x$, respectively.
	\end{remark}
	
	\begin{theorem}
	\label{thm1}
		Given the homogeneous memory-less vehicle meeting assumption \eqref{homo_meet}, the BETA policy is optimal in minimizing the deadline violation ratio in \eqref{pv}. 
	\end{theorem}
	\begin{IEEEproof}
	The proof is based on backwards induction on time. See Appendix \ref{app_thm1} for details.
	\end{IEEEproof}
	
	\begin{remark}
	A high-level intuition to explain the optimality of BETA policy is as follows. A stochastic coupling argument (the details are omitted for brevity) can be used to show that a necessary and sufficient condition for optimality is to avoid unnecessary service, which is defined as the event that a vehicle executes a task that has already been finished by vehicles before it. Based on this observation, the tasks should be assigned as balanced as possible, assuming the service occurs uniformly among tasks, to reduce the potential number of unnecessary services. The fundamental reason for the optimality of the BETA policy is that the tasks are treated equally, both from the deadline and the violation ratio perspectives. 
	\end{remark}
	
	\section{Performance Analysis for BETA Policy}
	\label{sec_pa}
	The BETA policy is proved to be optimal, as long as the homogeneous memory-less vehicle meeting assumption described in Section \ref{sec_sm} is upheld. The vehicle meeting process at the task-RSU is irrelevant, which can be observed by the proof of Theorem \ref{thm1}. Despite that the optimality of the BETA policy is shown, it is difficult to obtain a closed-form formula for the performance, i.e., concretely the deadline violation probability (average violation ratio) of the system. The difficulty stems from deriving the static distribution of a high-dimensional Markov chain described by the formulated MDP given the BETA policy. To address this issue and gain more insights into the performance, we derive a closed-form upper bound for the deadline violation probability of the eVCC system, which is shown to be quite tight, by noticing the fact that the BETA policy evenly replicates the tasks and therefore the performance of the BETA policy should be similar with an equivalent system where there is only one task with one $N$-th of the vehicles. The following theorem formalizes this finding.
	
	\begin{theorem}
		\label{thm2}
		The deadline violation probability of the system is upper bounded by 
		\begin{IEEEeqnarray}{rCl}
			\Pr\{t_\textrm{D}>D\} \le \exp \left\{-\frac{\lambda S}{N}\right(1&+&\frac{\lambda_1}{\lambda_2-\lambda_1}e^{-\lambda_2 D}  \nonumber\\
			&-&\left. \left.\frac{\lambda_2}{\lambda_2-\lambda_1}e^{-\lambda_1 D}\right)\right\},
		\end{IEEEeqnarray}
		where $\lambda_2=\mu B$ and $\lambda_1=\mu$.
	\end{theorem}
	
	\begin{IEEEproof}
	See Appendix \ref{app_thm2}.
	\end{IEEEproof}
	
	\begin{coro} 
		\label{coro_ray}
		\emph{(Short deadline regime)} When $D \to 0$, the task computation time lower bound derived in Theorem \ref{thm2}, i.e., $t_\textrm{D}$, follows the Rayleigh distribution, i.e., 	
		\begin{equation}
		\Pr\{t_\textrm{D}<D\} = 1-e^{-\frac{\lambda S}{2N}B\mu^2D^2},
		\end{equation}
		and the deadline violation probability upper bound is $1-\Pr\{t_\textrm{D}<D\}$ given deadline $D$. The mean computation delay is 	
		\begin{equation}
		\mathbb{E}[t_\textrm{D}]=\sqrt{\frac{N\pi}{2\lambda BS\mu^2}}.
		\end{equation}
	\end{coro}
	\begin{IEEEproof}
	See Appendix \ref{app_coro_ray}.
	\end{IEEEproof}
	\begin{remark}
	Since Corollary \ref{coro_ray} is the Taylor expansion approximation of Theorem \ref{thm2}, the short deadline regime is there when 
	\begin{equation}
	    \lambda_1 D \to 0,\textrm{ and }\lambda_2 D \to 0,
	\end{equation}
	which denote the average numbers of meetings between a vehicle and one and $B$ RSUs before deadline, respectively. The numbers are usually small in typical scenarios, making the approximation accurate in most cases. Therefore, Corollary \ref{coro_ray} is used in the following section for ease of exposition.
	\end{remark}
	
	\section{Implications on Road Traffic: Optimum Vehicle Density for eVCC}
	\label{sec_imp}
	\begin{figure*}[!t]
    \centering
    \includegraphics[width=0.75\textwidth]{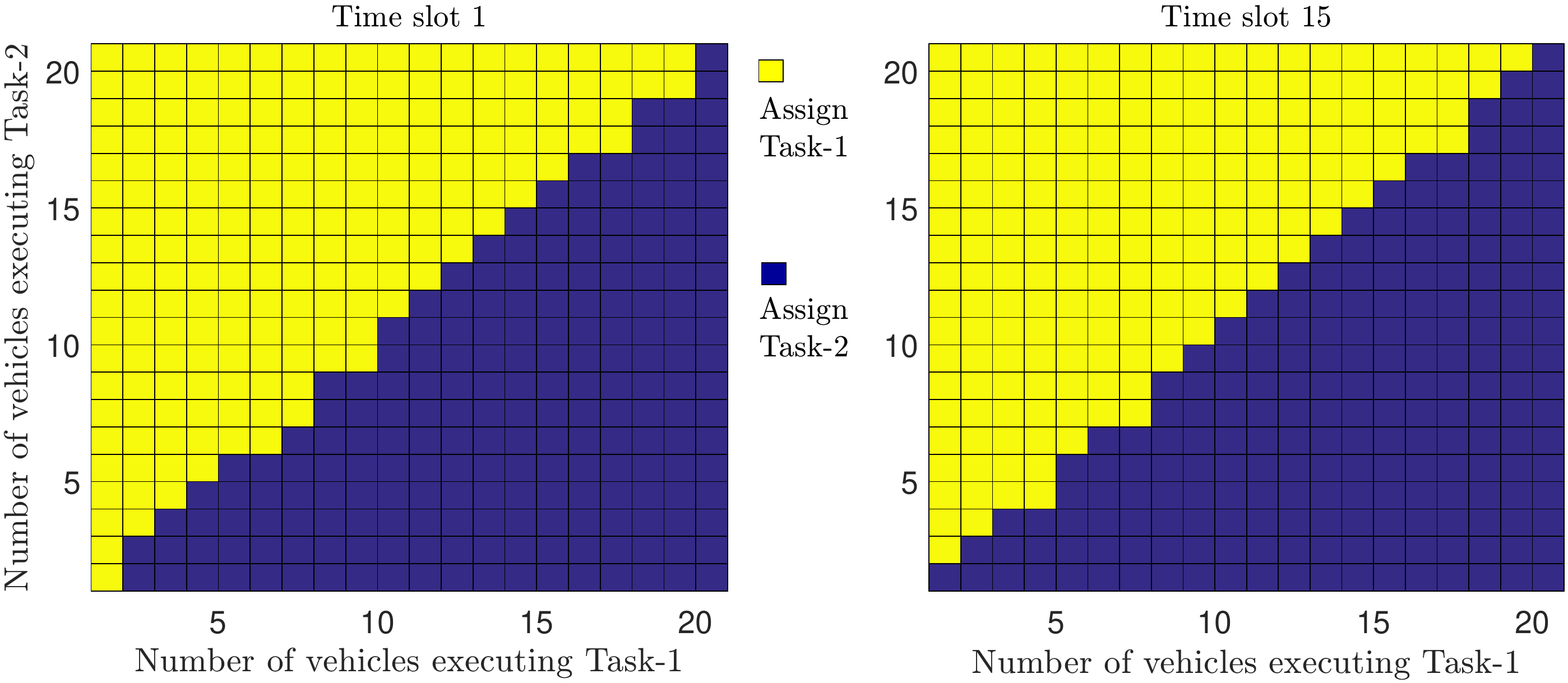}
    \caption{The optimal task assignment policy given by solving MDP-$1$. The total number of vehicles is $M=20$, the number of RSUs is $B=5$, the delay deadline is $D=20$, and only two tasks are considered.}
    \label{Fig_mdp_policy}
    \end{figure*}
    \begin{figure}[!t]
		\centering
		\includegraphics[width=0.48\textwidth]{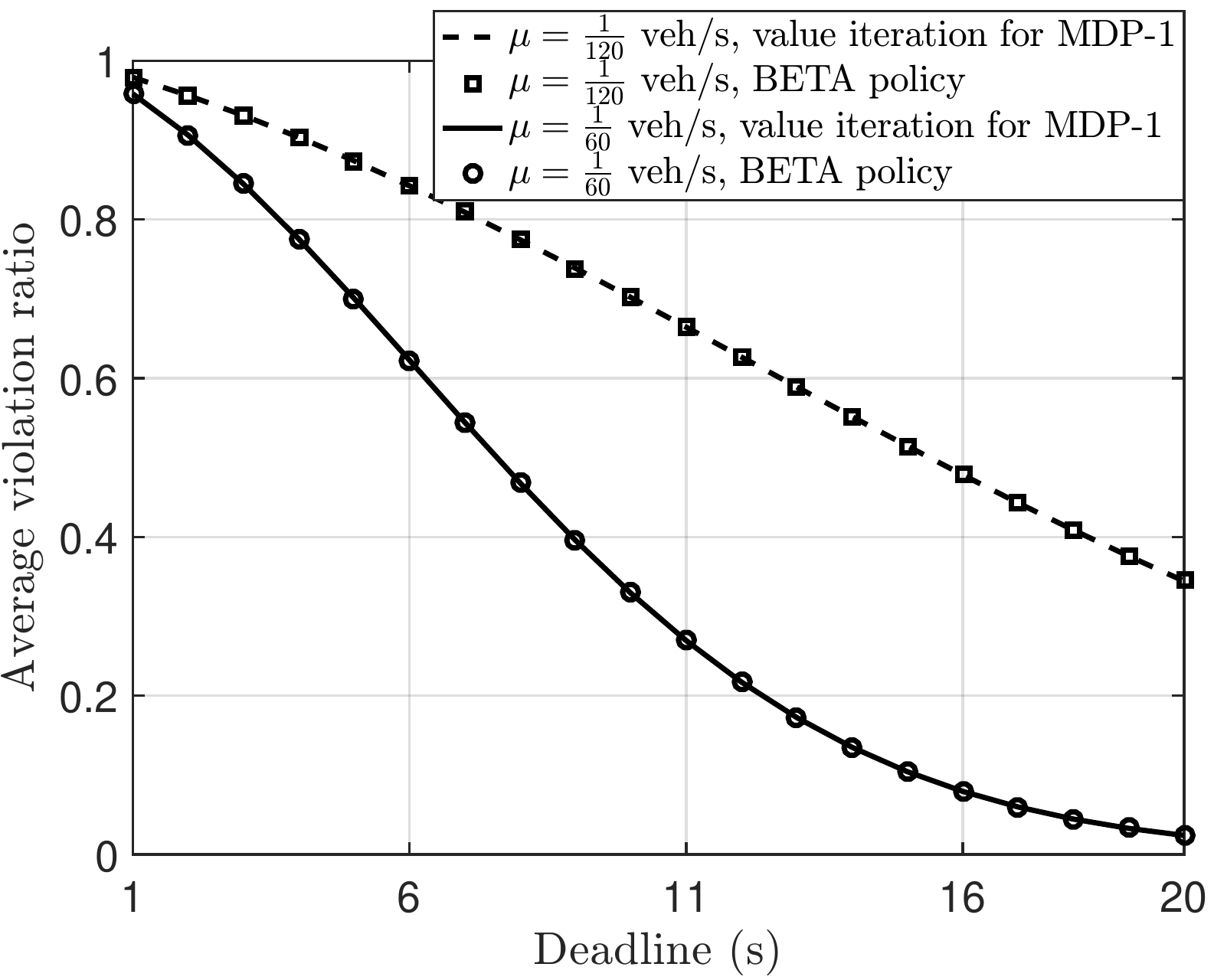}
		\caption{The optimal average violation ratio given by value iteration to solve MDP-$1$, in comparison with the achieved average violation ratio by the BETA policy. The total number of vehicles is $M=20$, the number of RSUs is $B=5$ and only two tasks are considered.}
		\label{Fig_mdp_value}
	\end{figure}
	\begin{figure*}[!t]
        \centering
        \subfigure[$L = 30$~veh/km.]{
        \includegraphics[width=0.235\textwidth]{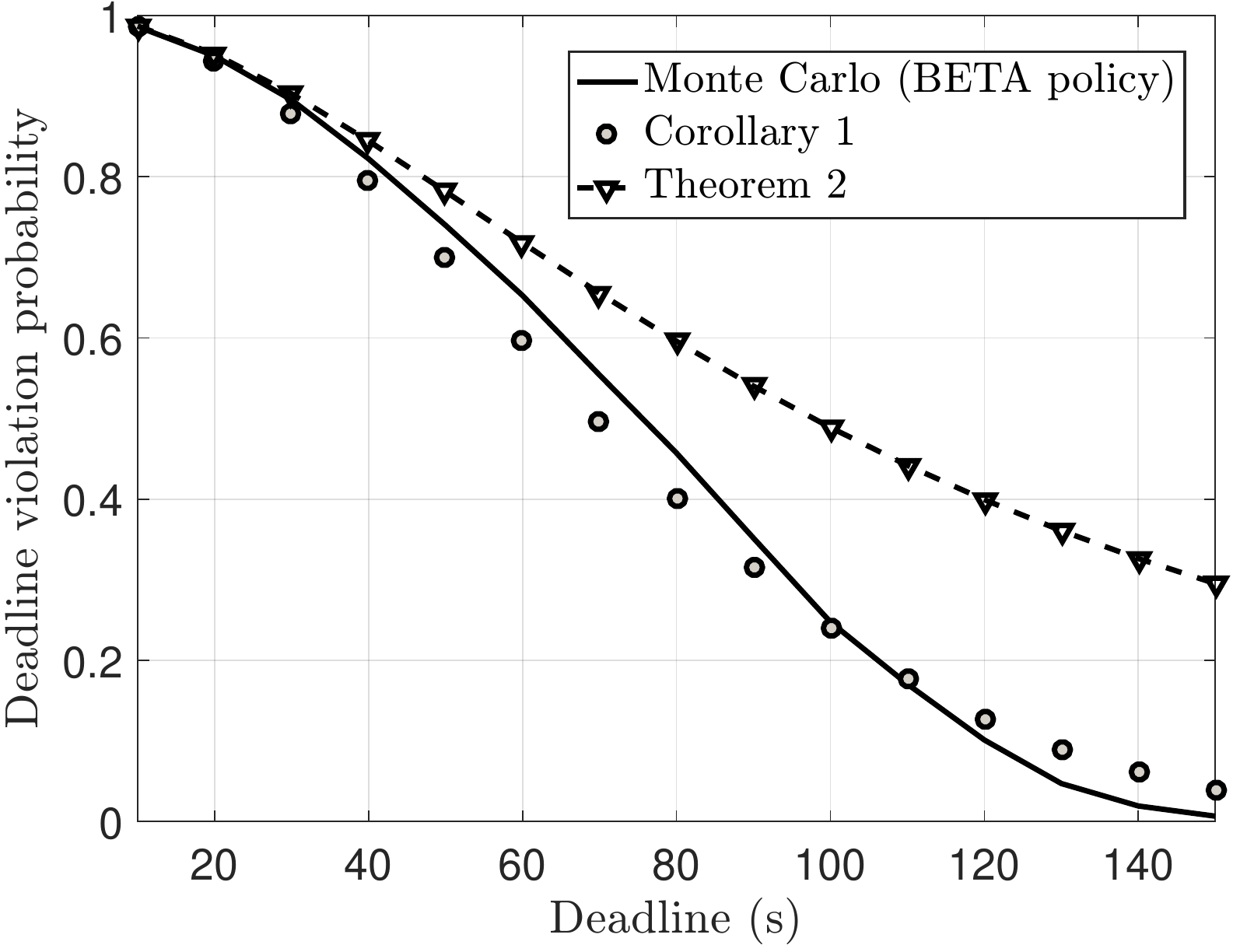}}
        \subfigure[$L = 60$~veh/km.]{
        \includegraphics[width=0.235\textwidth]{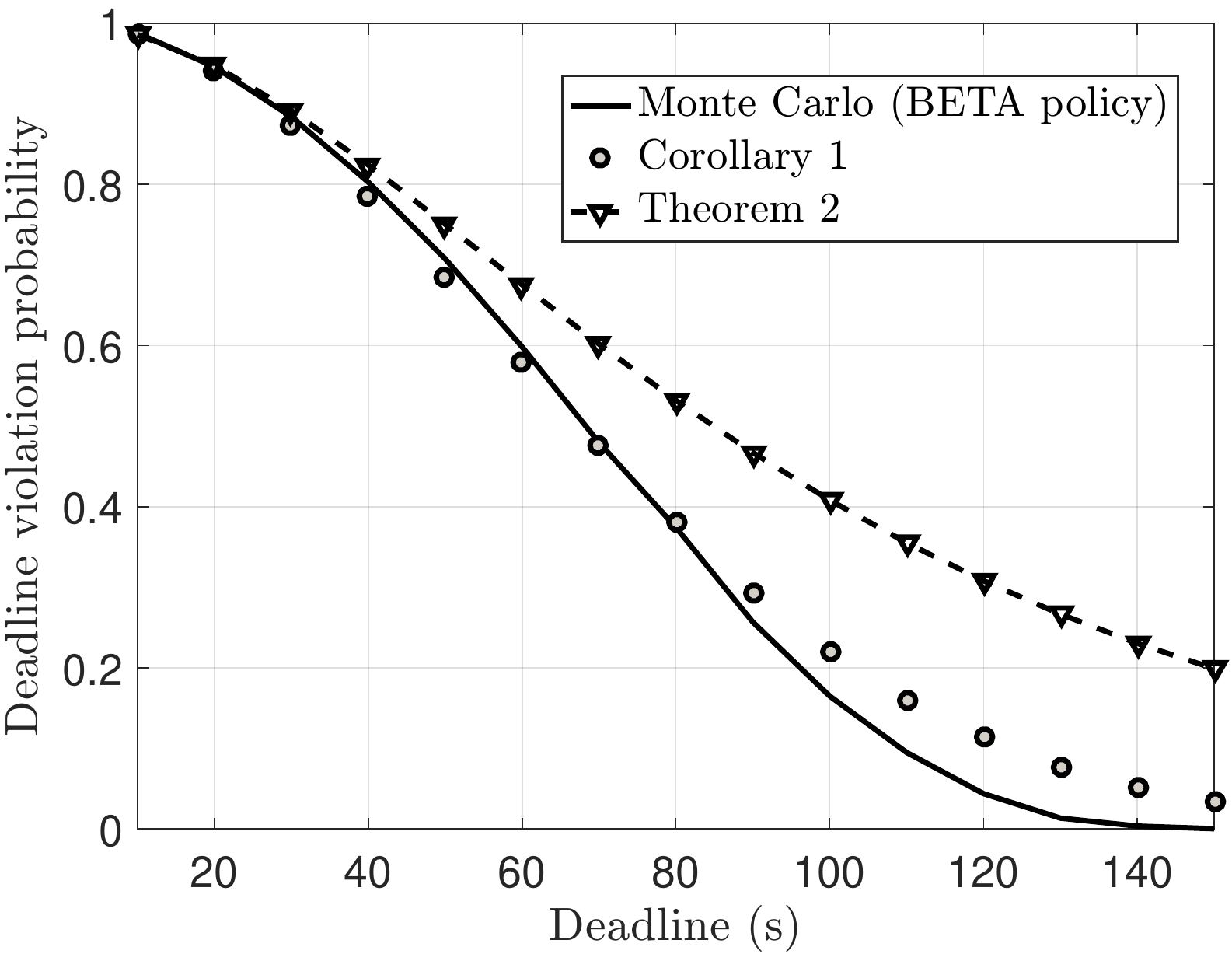}}
        \subfigure[$L = 90$~veh/km.]{
        \includegraphics[width=0.235\textwidth]{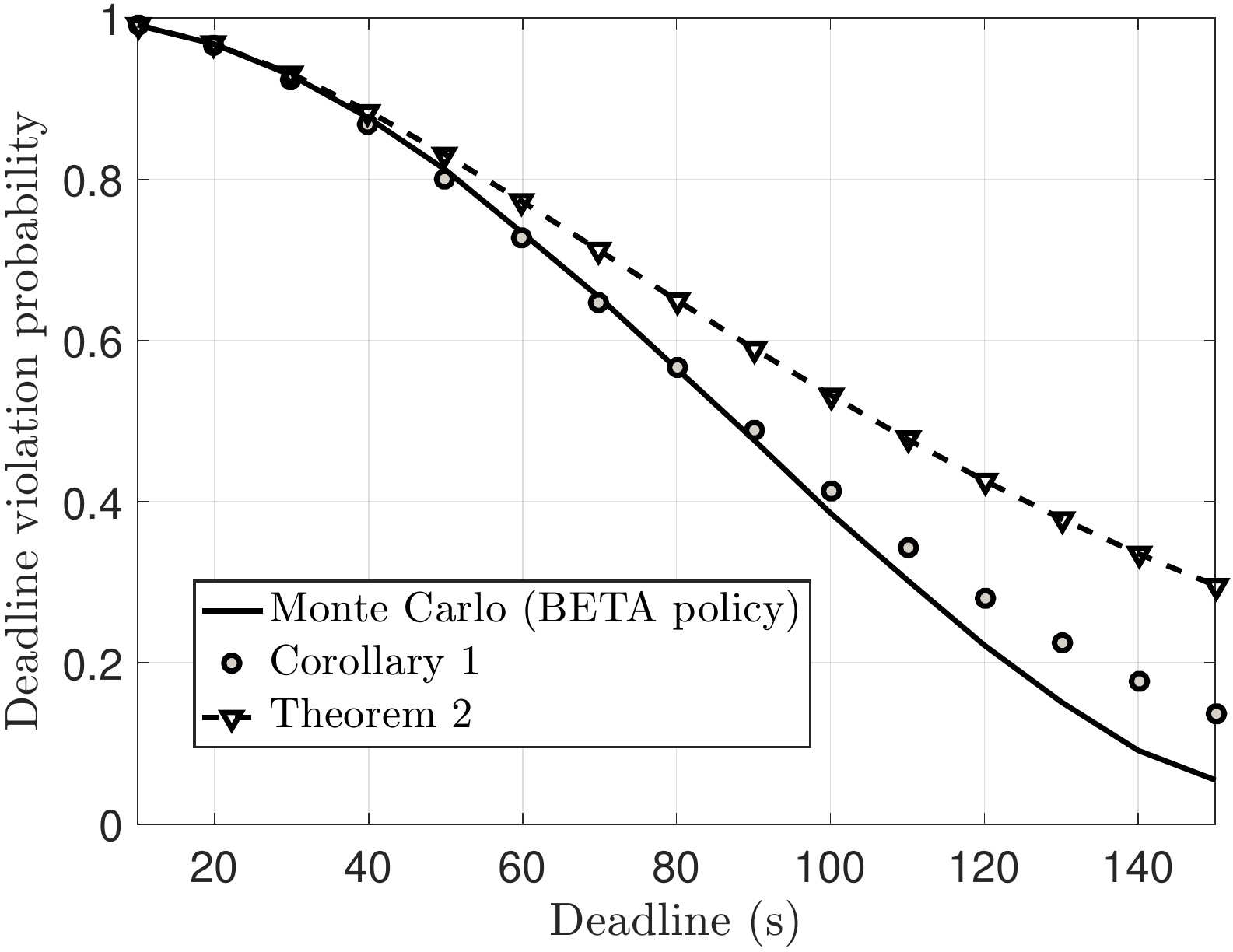}}
        \subfigure[$L = 120$~veh/km.]{
        \includegraphics[width=0.235\textwidth]{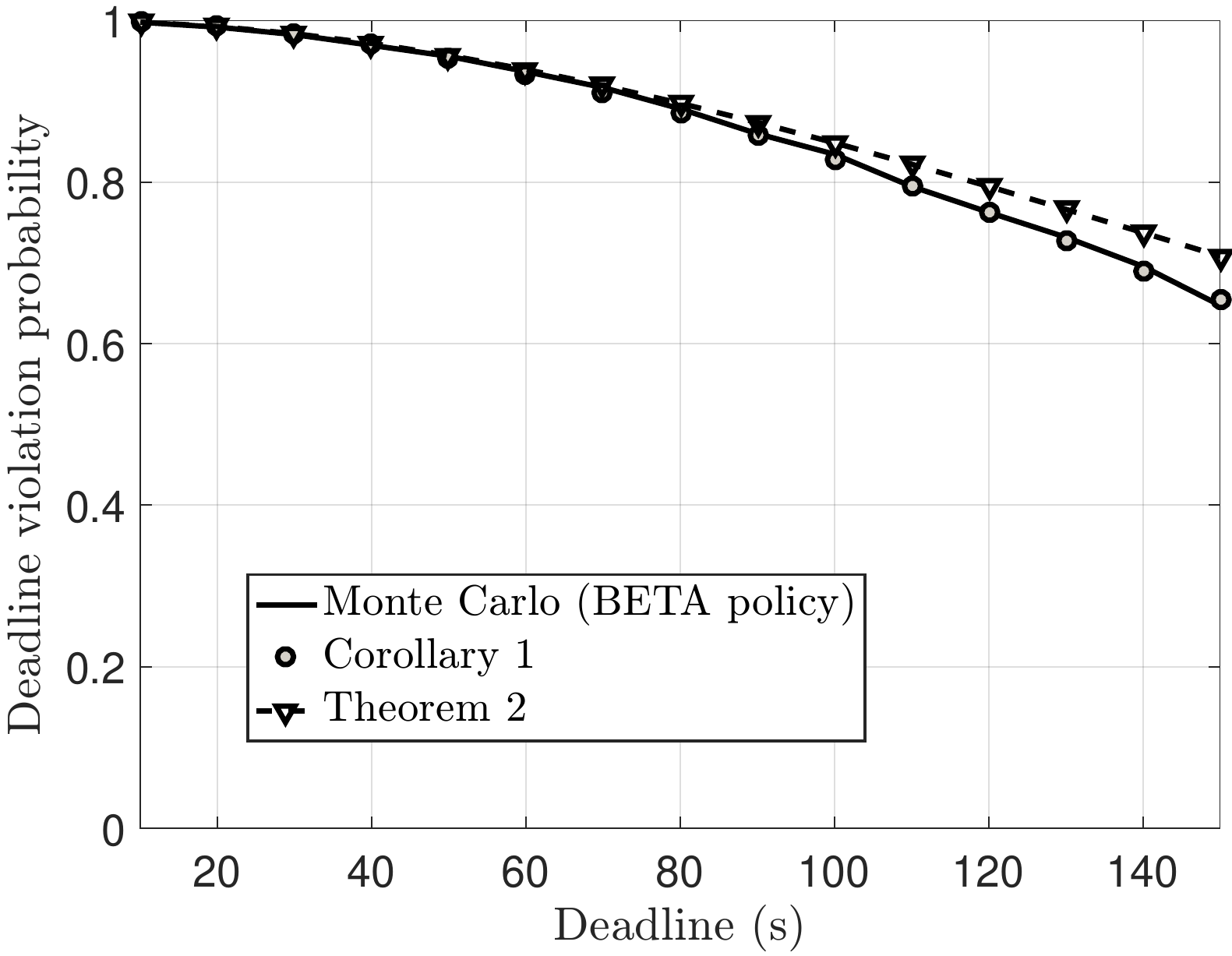}}
        \caption{The optimal average violation ratio versus the deadline given by Monte Carlo simulations based on the BETA policy, in comparison with the violation probability upper bound given in Theorem \ref{thm2} and Corollary \ref{coro_ray}. The number of RSUs is $B=10$, the road length is $10$~km, the number of tasks is $50$.}
        \label{Fig_p_d}
    \end{figure*}
    \begin{figure*}[!t]
        \centering
        \subfigure[$L = 30$~veh/km.]{
        \includegraphics[width=0.235\textwidth]{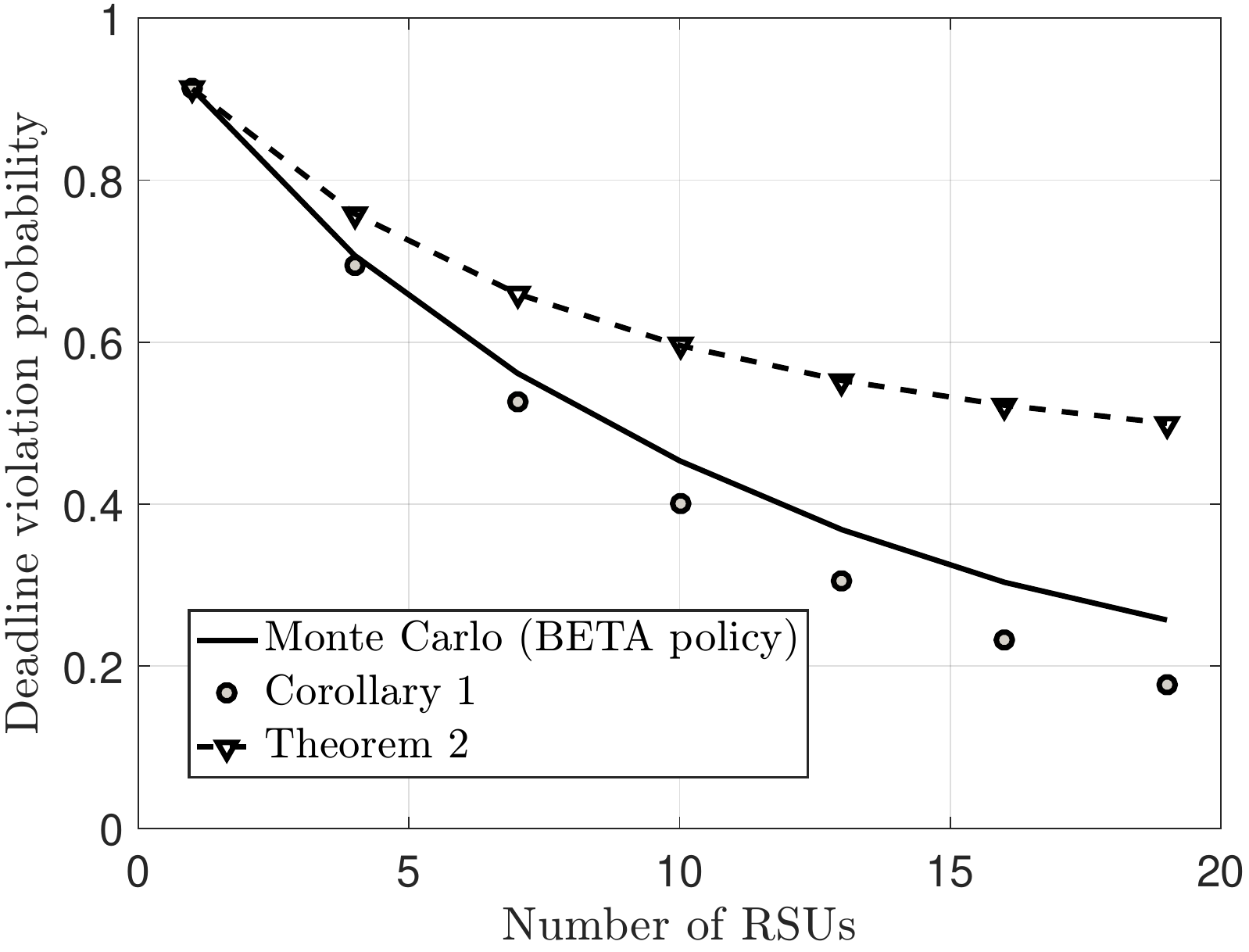}}
        \subfigure[$L = 60$~veh/km.]{
        \includegraphics[width=0.235\textwidth]{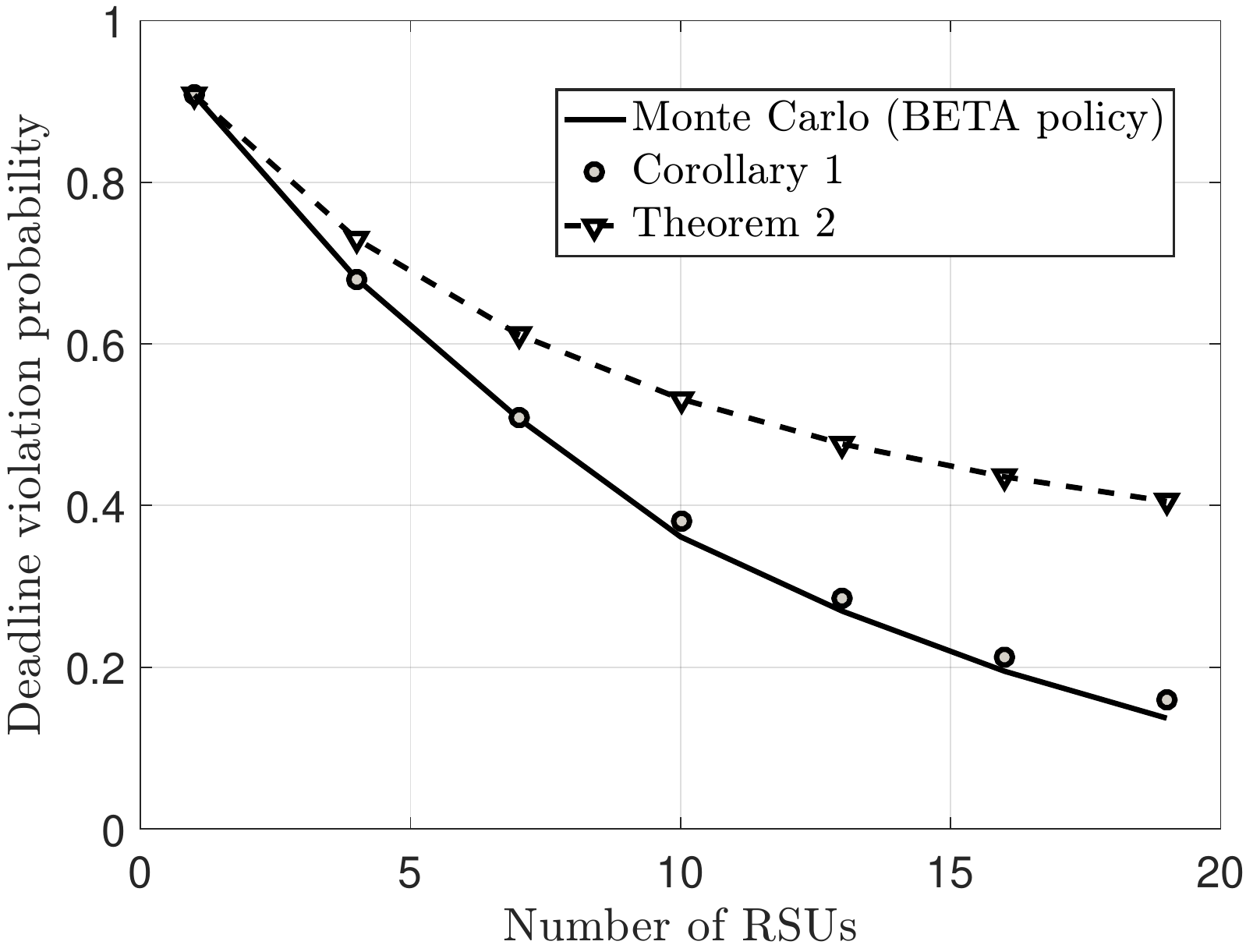}}
        \subfigure[$L = 90$~veh/km.]{
        \includegraphics[width=0.235\textwidth]{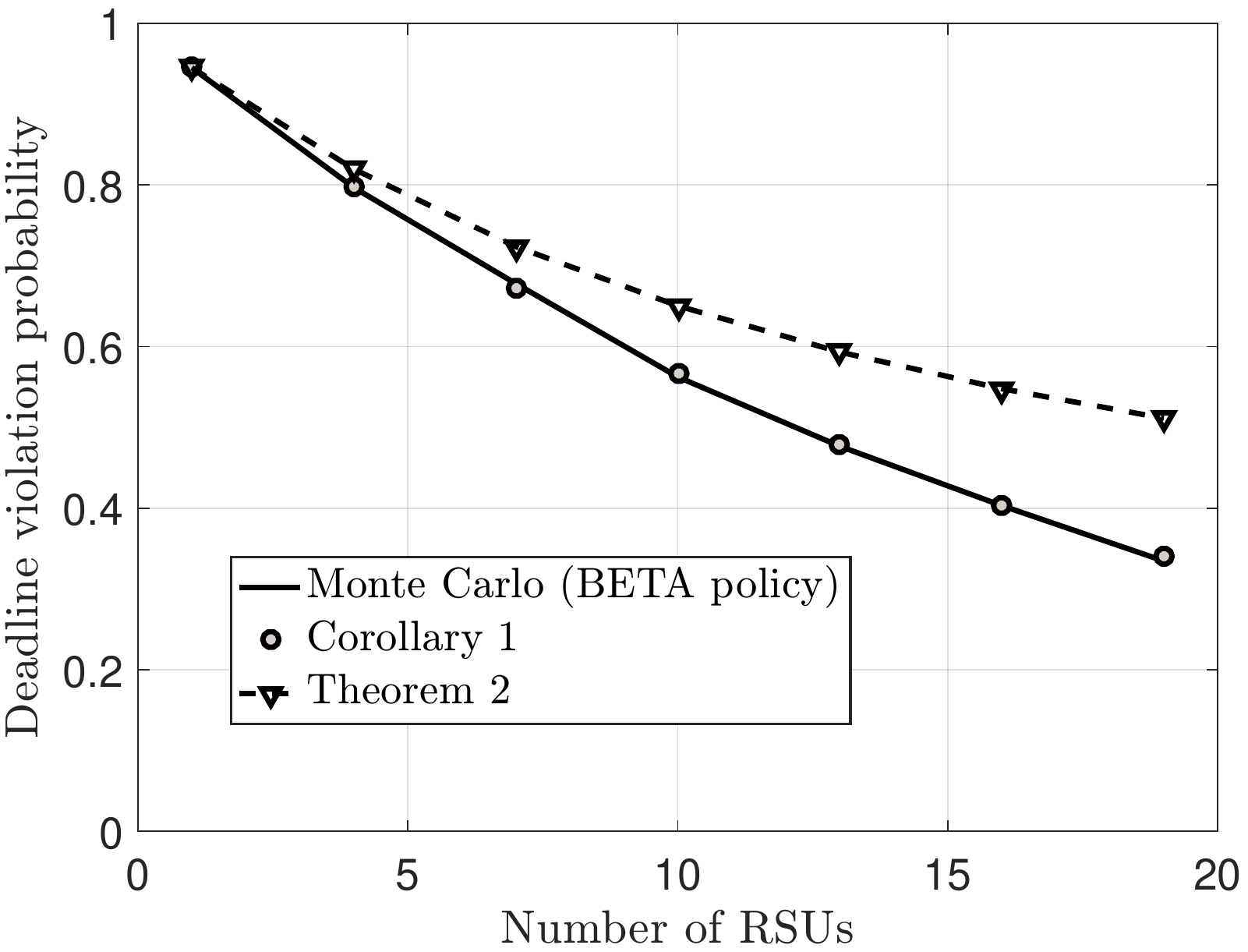}}
        \subfigure[$L = 120$~veh/km.]{
        \includegraphics[width=0.235\textwidth]{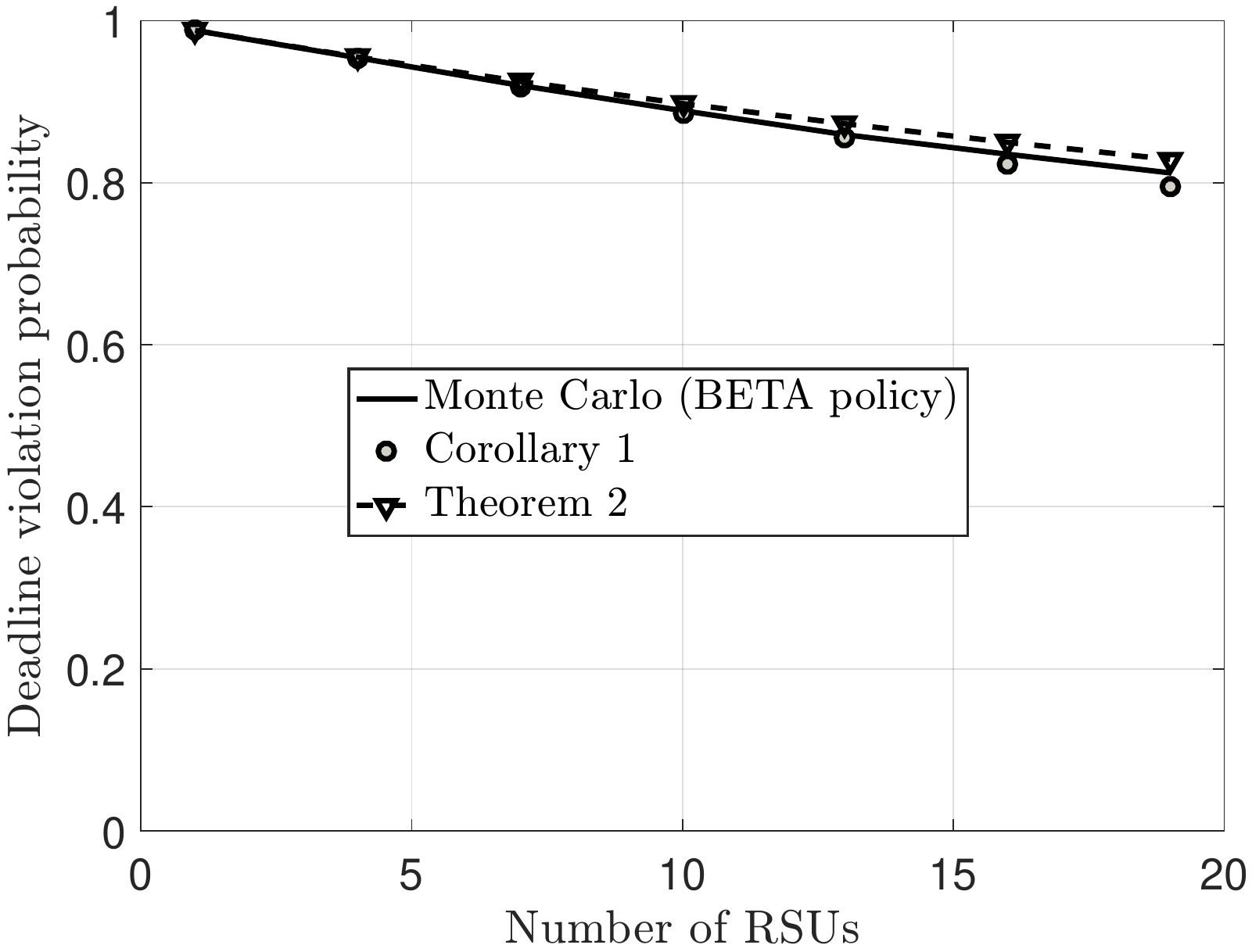}}
        \caption{The optimal average violation ratio versus the number of RSUs on the road given by Monte Carlo simulations based on the BETA policy, in comparison with the violation probability upper bound given in Theorem \ref{thm2} and Corollary \ref{coro_ray}. The road length is $10$~km, the number of tasks is $50$, the deadline is $80$~s.}
        \label{Fig_p_b}
    \end{figure*}
    \begin{figure*}[!t]
        \centering
        \subfigure[$L = 30$~veh/km.]{
        \includegraphics[width=0.235\textwidth]{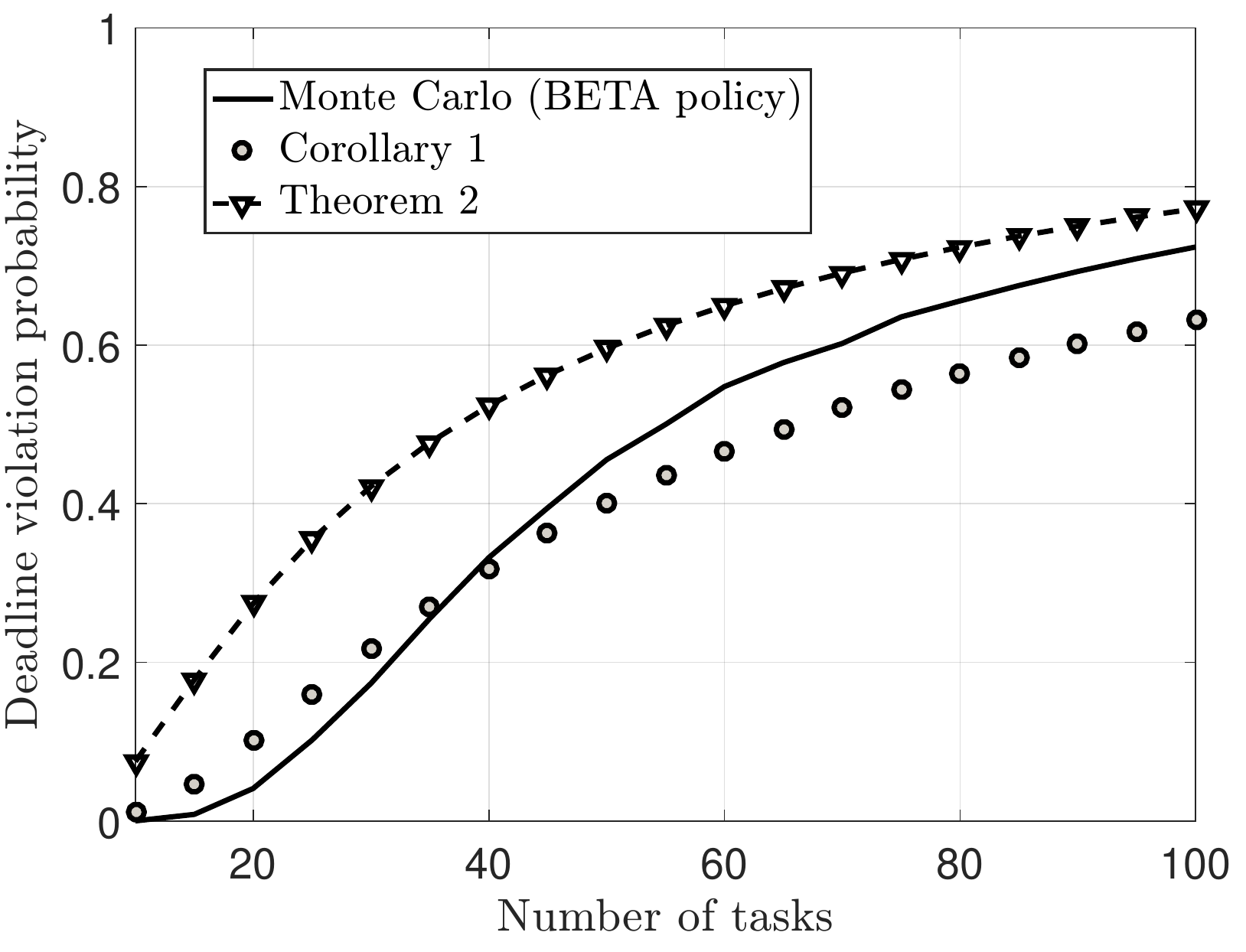}}
        \subfigure[$L = 60$~veh/km.]{
        \includegraphics[width=0.235\textwidth]{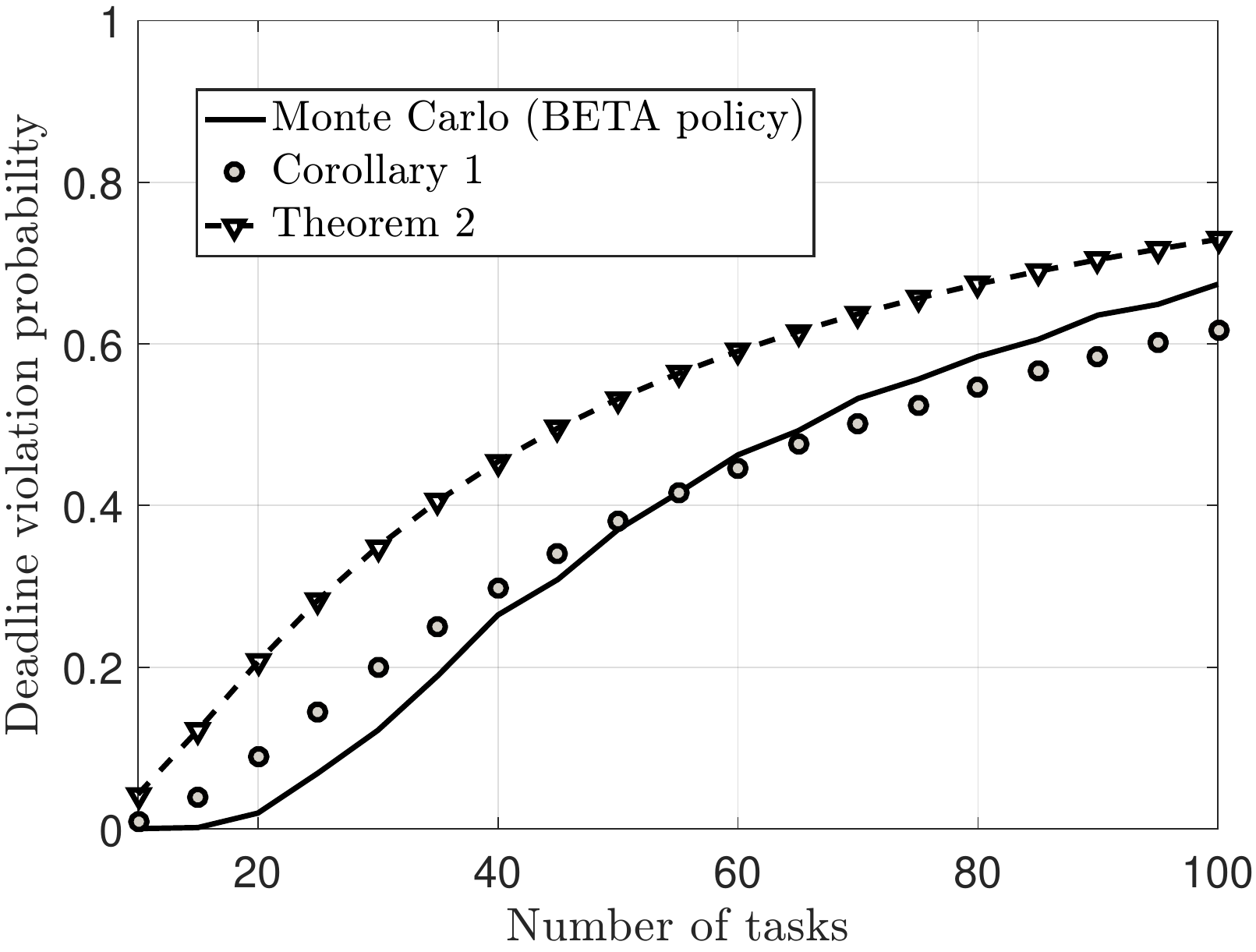}}
        \subfigure[$L = 90$~veh/km.]{
        \includegraphics[width=0.235\textwidth]{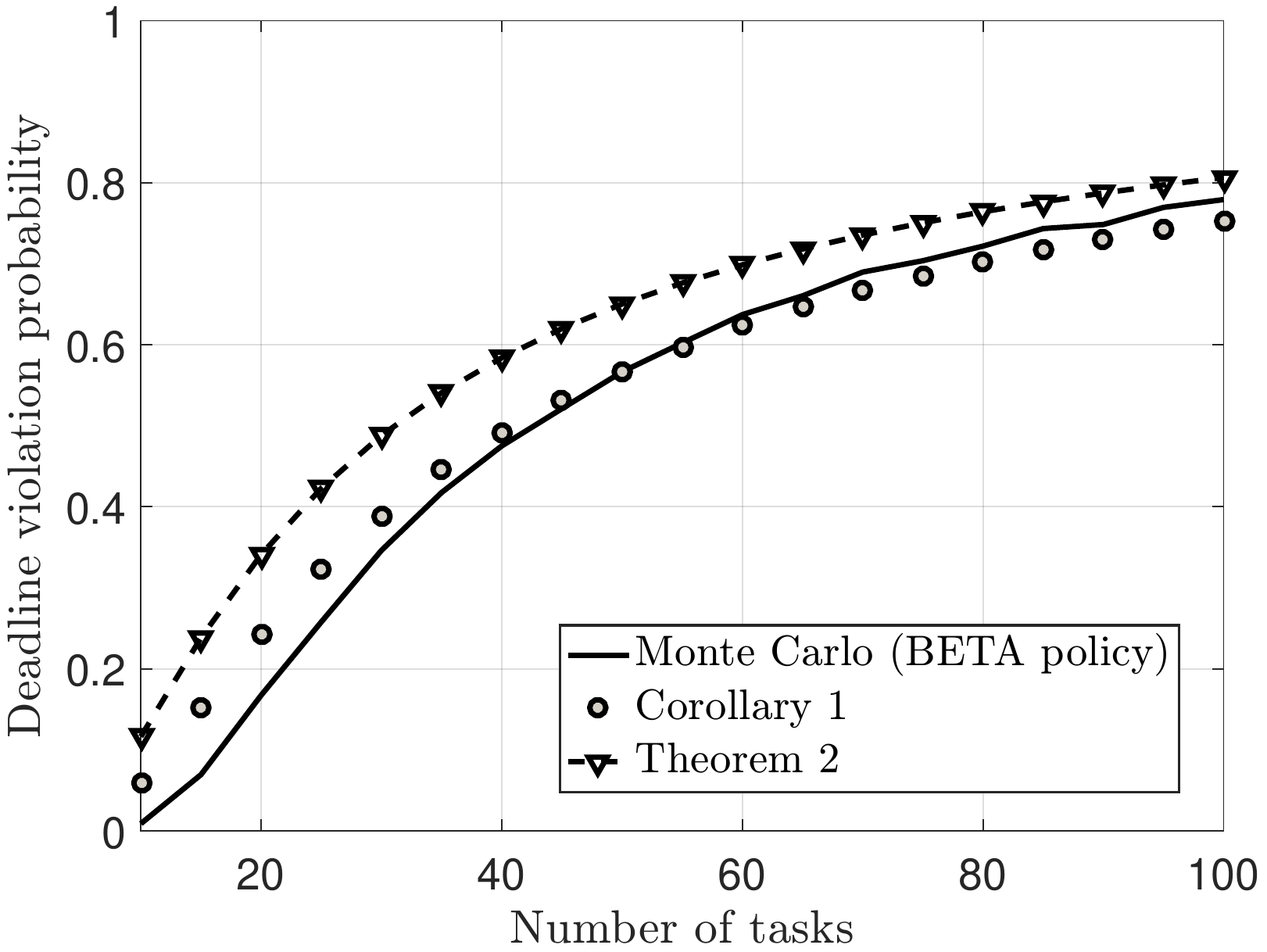}}
        \subfigure[$L = 120$~veh/km.]{
        \includegraphics[width=0.235\textwidth]{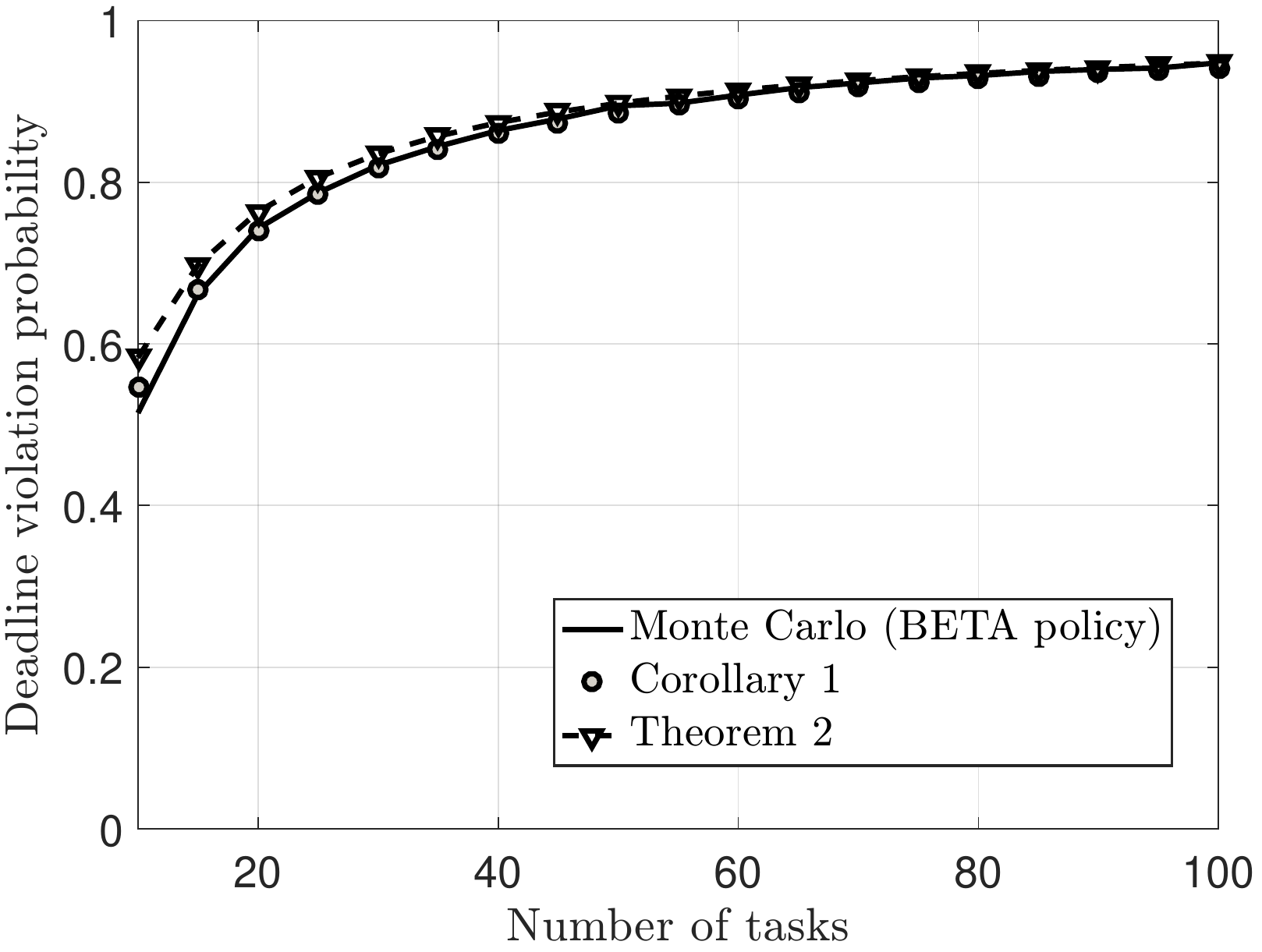}}
        \caption{The optimal average violation ratio versus the number of tasks given by Monte Carlo simulations based on the BETA policy, in comparison with the violation probability upper bound given in Theorem \ref{thm2} and Corollary \ref{coro_ray}. The road length is $10$~km, the number of RSUs is $10$, the deadline is $80$~s.}
        \label{Fig_p_n}
    \end{figure*}
    \begin{figure}[!t]
        \centering
        \includegraphics[width=0.48\textwidth]{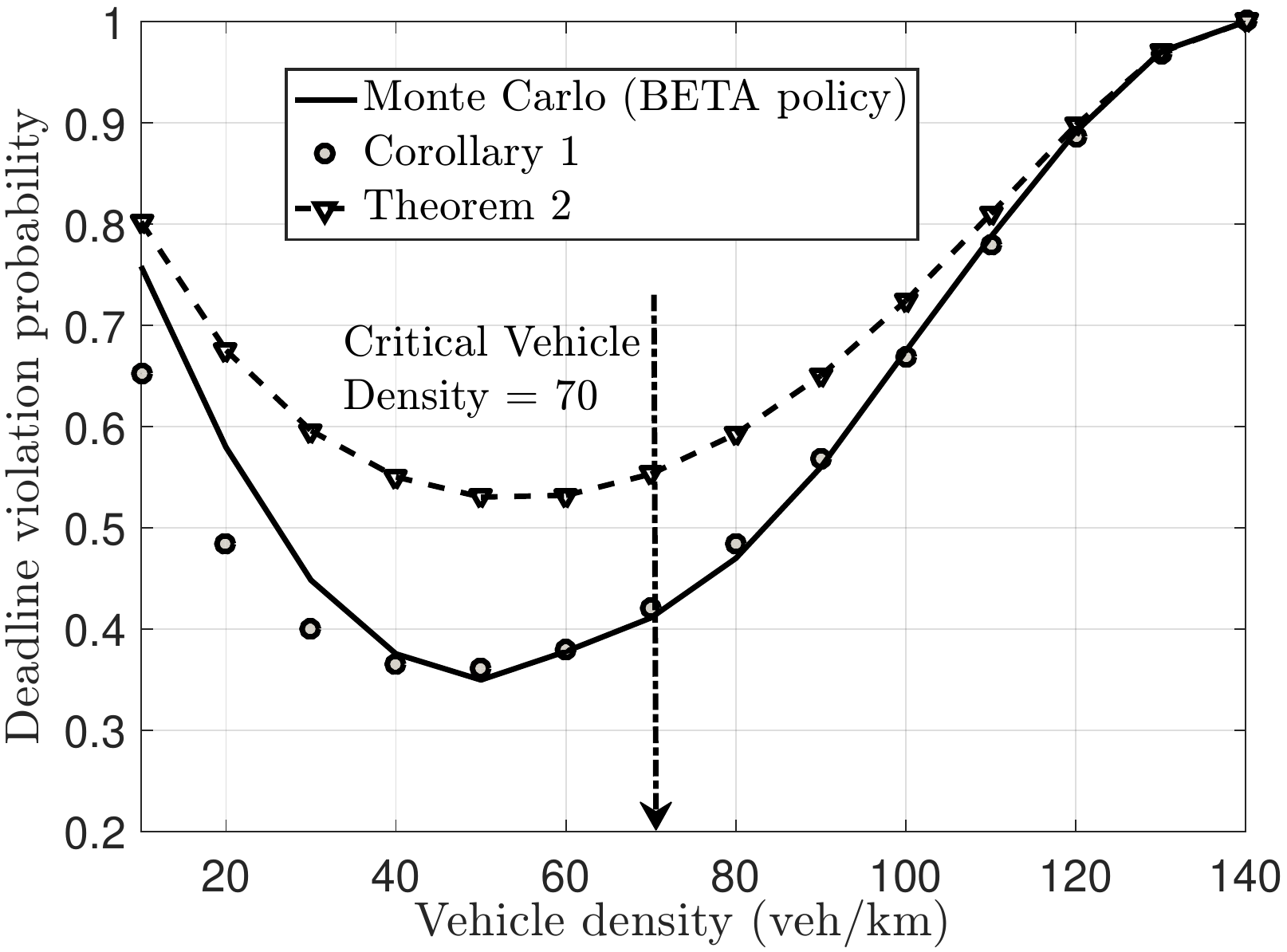}
        \caption{The optimal average violation ratio versus the vehicle density on the road given by Monte Carlo simulations based on the BETA policy, in comparison with the violation probability upper bound given in Theorem \ref{thm2} and Corollary \ref{coro_ray}. The road length is $10$~km, the number of RSUs is $10$, the number of tasks is $50$, the deadline is $80$~s.}
        \label{Fig_p_l}
    \end{figure}
    \begin{figure}[!t]
        \centering
        \includegraphics[width=0.48\textwidth]{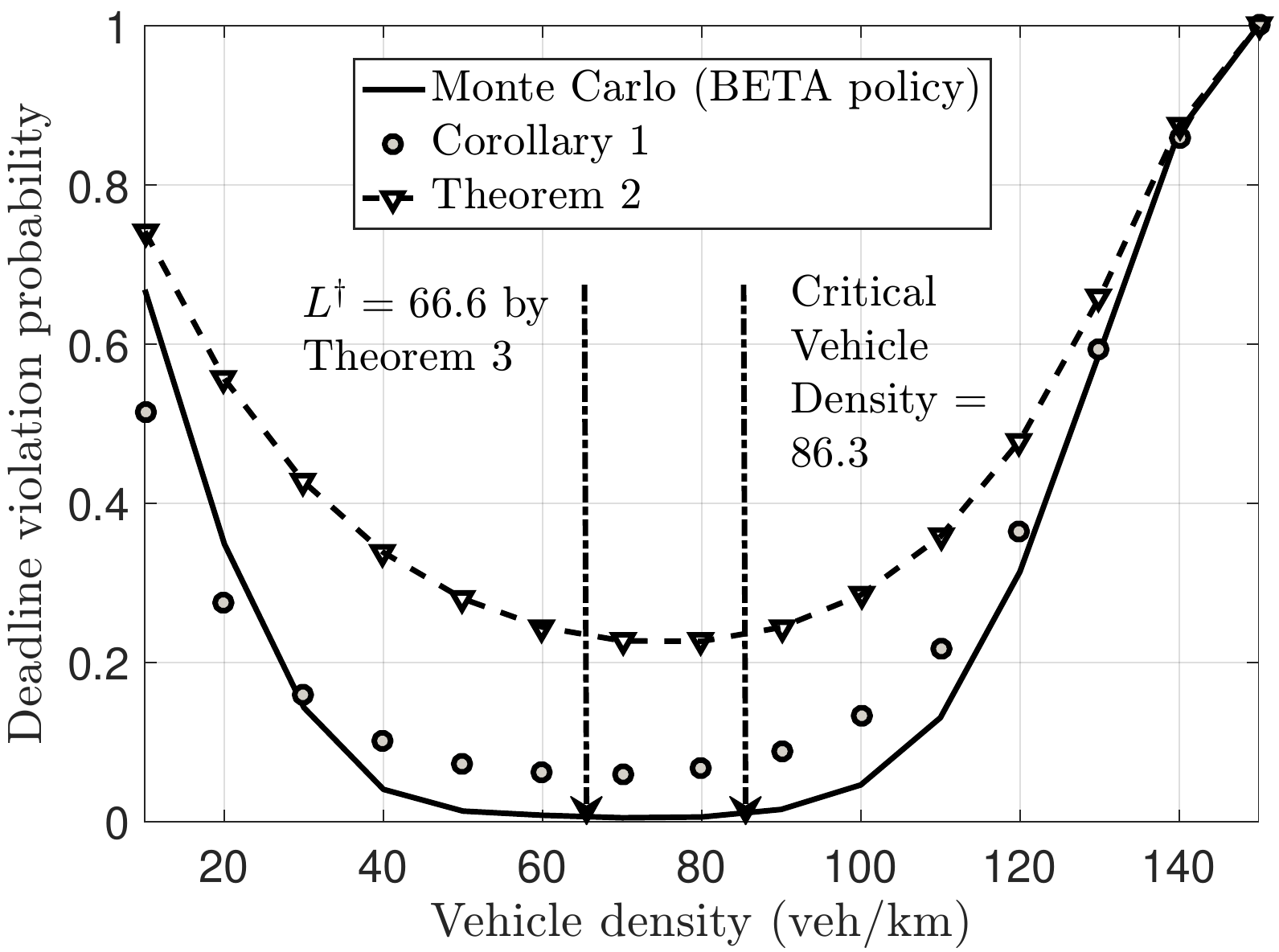}
        \caption{Based on measurement validated model in \cite{toledo04} \cite{pipes67}, the optimal average violation ratio versus the vehicle density on the road given by Monte Carlo simulations based on the BETA policy, in comparison with the violation probability upper bound given in Theorem \ref{thm2} and Corollary \ref{coro_ray}. The road length is $10$~km, the number of RSUs is $10$, the number of tasks is $50$, the deadline is $80$~s.}
        \label{Fig_p_l2}
    \end{figure}
    \begin{figure}[!t]
        \centering
        \includegraphics[width=0.48\textwidth]{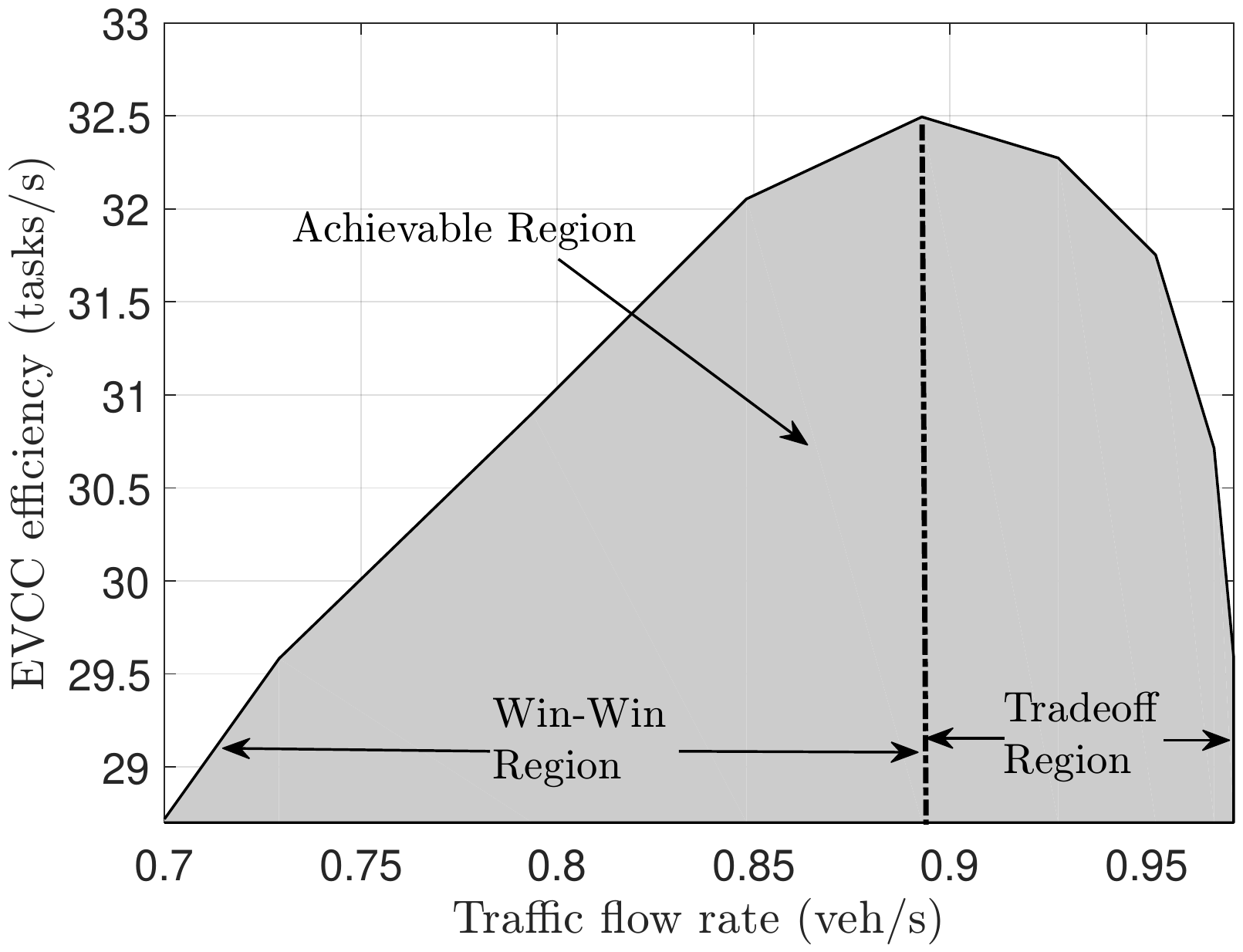}
        \caption{The tradeoff between eVCC efficiency and traffic efficiency. The road length is $10$~km, the number of RSUs is $10$, the number of tasks is $50$, the deadline is $80$~s.}
        \label{Fig_ct}
    \end{figure}
	In this section, some insights are given by combining the performance analysis of the optimal BETA policy and the traffic engineering theory \cite{rose11}. It is observed that the average vehicle speed decreases with the vehicle density in the real world due to limited human reaction time \cite{rose11}. Therefore, to minimize the deadline violation probability, the impacts of vehicle density and vehicle speed, which reflect on the number of SVs and the output collection rate respectively, need to be jointly considered to find the optimum tradeoff. 
	
	Based on the traffic theory, there are three macroscopic parameters that describe the traffic flow state on the road: the average (space mean) speed of traffic stream $V$, the rate of flow $F$, and the vehicle density $L$. With the unified units, the fundamental equation of the speed-flow-density relationships is 
	\begin{equation}
	F = VL.
	\end{equation}
	The general relationship of the basic parameters can be well illustrated by the approximate linear relationship between the speed of vehicles and the vehicle density \cite{yousef08}, i.e.,
	\begin{equation}
	\label{VD}
	V(L) = V_\textrm{max}\left(1-\frac{L}{L_\textrm{max}}\right),
	\end{equation}
	where $V_\textrm{max}$ is the speed limit, and $L_\textrm{max}$ is the traffic jam vehicle density. Later in the section, the linear relationship will be generalized. In the traffic theory, there is a concept named \emph{critical vehicle density}, which is defined as the optimum vehicle density given a stationary traffic flow that maximizes the traffic efficiency, i.e., the traffic flow rate. It is straightforward to see that the critical vehicle density to maximize the vehicle flow rate given \eqref{VD} is 
	\begin{equation}
	\label{two}
	L^* = \frac{L_\textrm{max}}{2}.
	\end{equation}
	Essentially, it states the simple fact that when there is too much traffic wherein the vehicles cannot move or when there is too little traffic wherein the vehicles are too few, the traffic flow throughput is not satisfactory. 
	
	The question is whether the critical vehicle density is also optimal for the eVCC system. Towards this end, we have the following theorem.
	
	\begin{theorem}
	\label{thm3}
	When the speed-density relationship, i.e., $V(L)$, satisfies the following conditions, the optimum vehicle density for eVCC system $L^\dag$ is smaller than the critical vehicle density $L^*$.
	\begin{enumerate}
	    \item 
	    $V(L)$ is monotonically non-increasing, and $V(L) \ge 0$ for $0 \le L \le L_\textrm{max}$.
	    \item
	    $V(L)$ is a concave function of $L$, or it satisfies 
	    \begin{equation}
	    \label{v_condition2}
	        V(L) = G - \sum_{k=1}^{K} c_k L^{\alpha_k},
	    \end{equation}
	    where $\alpha_k \in (0,+\infty)\cup(-1,-\frac{1}{2})$, $c_k \ge 0$, $\forall k$ and $G \ge 0$.
	\end{enumerate}
	\end{theorem}
	\begin{IEEEproof}
	See Appendix \ref{app_thm3}.
	\end{IEEEproof}
	
	\begin{remark}
	Note that the first condition in Theorem \ref{thm3} is almost always met in practice, due to the fact that people would drive more slowly with more vehicles on the road \cite{van95}. The second condition allows several well-recognized speed-density models to be applicable, e.g., in \cite[Chapter 12]{drew68} and \cite{pipes67}.
	\end{remark}
	
	\begin{remark}
	Theorem \ref{thm3} establishes that under most circumstances, the optimum vehicle density for the eVCC system is smaller than the critical vehicle density. Therefore, the corresponding optimum speed comparison is the opposite, which indicates that vehicle mobility actually benefits the eVCC system more than the traditional traffic system.
	\end{remark}
	
	\begin{coro}
	\label{prop1}
		In the short deadline regime, if the linear speed-density relationship is satisfied, the optimum vehicle density to minimize the average deadline violation probability is
		\begin{equation}
		L^\dag = \frac{L_\textrm{max}}{3}.
		\end{equation}
	\end{coro}
	
	\begin{IEEEproof}
		See Appendix \ref{app_coro2}.
	\end{IEEEproof}
	\begin{remark}
	This result corresponds to the critical vehicle density with linear speed-density model in \eqref{two}. It demonstrates, in closed-form, the comparisons between the optimal vehicle densities in VCC systems and traffic systems.
	\end{remark}
	\subsection{Asymptotic Analysis}
	\label{sec_aa}
	\subsubsection{Large City Regime}
	In the large city regime, the length of the road is approaching infinity, i.e., $S \to \infty$. Since the number of available vehicles is increased whereas the distance of travelling is also increased, it is interesting to investigate the performance in the regime. Based on Theorem \ref{thm2}, it follows that when $S \to \infty$, $\mu \to 0$, then the deadline violation probability approaches the following based on the linear speed-density relationship \eqref{VD},
	\begin{IEEEeqnarray}{rCl}
	Pr(t_{\textrm{D}}>D) &\xrightarrow{{S \to \infty}}& \exp \left(-\frac{L S}{2N}B\mu^2D^2\right) \nonumber\\
	&\xrightarrow{{S \to \infty}}& \exp\left({-\frac{L V_\textrm{max}^2 B}{2SN}\left(1-\frac{L}{L_\textrm{max}}\right)^2D^2}\right). \nonumber\\
	\end{IEEEeqnarray}
	It can be observed that as long as the RSU density, i.e., $B/S$, and the vehicle density remains unchanged, the performance is irrelevant with the road length. 
	\subsubsection{High RSU Density Regime}
	When the RSU density goes to infinity, i.e., $B/S \to \infty$, combining with Theorem \ref{thm2}, we obtain
	\begin{equation}
	\label{largeB}
	    Pr(t_{\textrm{D}}>D) \xrightarrow{\frac{B}{S} \to \infty} \exp{\left(-\frac{L S}{N} \left(1-e^{-\mu D}\right)\right)}.
	\end{equation}
	If we continue to assume the short deadline regime as in Corollary \ref{coro_ray}, then \eqref{largeB} becomes
	\begin{equation}
	\label{largeBsmallD}
	    Pr(t_{\textrm{D}}>D) \xrightarrow{\frac{B}{S} \to \infty,\, D \to 0} \exp{\left(-\frac{L V_\textrm{max}}{N} \left(1-\frac{L}{L_\textrm{max}}\right) D\right)}.
	\end{equation}
	It turns out that the optimum vehicle density in this regime is identical with the critical vehicle density which maximizes the traffic flow rate. This is explained by the fact that when RSUs are everywhere, the tasks would be immediately finished after they are offloaded, and hence the task computation time is determined by the vehicle meeting rate at the task-RSU, which is equivalent with the traffic flow rate assuming homogeneous vehicle meetings.
	
	\section{Numerical Results}
	\label{sec_nr}
	In this section, numerical results based on computer simulation are given to demonstrate the optimality of the BETA policy, to validate the tightness of the performance upper bound given in Theorem \ref{thm2} and Corollary \ref{coro_ray}, and to show the impacts of several key system parameters. The vehicle meeting rate $\mu$ is given by \eqref{VD}, where the speed limit is $V_\textrm{max}=100$~km/h and the traffic jam vehicle density is $L_\textrm{max} = 140$~veh/km assuming two lanes \cite{rose11}. 
    
    Fig. \ref{Fig_mdp_policy} and \ref{Fig_mdp_value} are shown to demonstrate the optimality of the BETA policy. The optimal policy is obtained by value iteration to solve MDP-$1$. It is observed in Fig. \ref{Fig_mdp_policy} that the optimal policy has the same structure with the BETA policy, which always assigns the task with fewer replicas. It is worth mentioning that in general the optimal policy for a finite-horizon MDP is not stationary, in the sense that it may change as the time goes on. However, it is shown that the BETA policy is optimal at any time slot (time slot $1$ and $15$ are shown). Also note that the ties are broken arbitrarily, and therefore the policy on the diagonal can go either way. Fig. \ref{Fig_mdp_value} shows that optimum value of the initial state equals the average violation ratio obtained by simulating the BETA policy, with two distinct departure rates. 
    
    Fig. \ref{Fig_p_d}-\ref{Fig_p_l2} are shown to reflect the impact of several key parameters, and to compare the simulation results with the closed-form expressions in Theorem \ref{thm2} and Corollary \ref{coro_ray}. The simulation results are obtained by running the BETA policy for $1000$ iterations and calculating the average, where in each iteration the number of vehicles is generated based on the Poisson distribution as assumed in Section \ref{sec_sm}. In general, it is shown that the deadline violation probability performance can be well characterized by the closed-form upper bounds given in Theorem \ref{thm2} and Corollary \ref{coro_ray}. In particular, the result in Corollary \ref{coro_ray} is very close to the simulation results. The optimum vehicle density given in Proposition \ref{prop1}, which minimizes the deadline violation probability is validated in Fig. \ref{Fig_p_l}. It is shown that the promptness of the eVCC system with the optimum vehicle density by Proposition \ref{prop1} is significantly better than the traditional critical vehicle density in the traffic engineering theory, demonstrating that vehicle mobility benefits the eVCC system more compared with the traffic system. It also indicates that we need to rethink the desired traffic conditions in the future when vehicular network based applications such as eVCC, are widely implemented. In Fig. \ref{Fig_p_l2}, the performance under a widely used speed-density model, which is validated based on real traffic measurements collected by using $3$-h a.m. peak sensor data from a section of the M$27$ freeway in Southampton England \cite{toledo04}. Based on the model, the speed-density relationship is characterized by
    \begin{equation}
        V = 116.4\left[1-\left(\frac{L}{149.797}\right)^{1.964}\right],
    \end{equation}
    which satisfies the second condition in Theorem \ref{thm3}.
    
    \subsection{Tradeoff between eVCC Efficiency and Traffic Efficiency}
    Define the eVCC efficiency as the average number of executed tasks in unit time, i.e.,
	\begin{equation}
	    \eta_\textrm{CE} = \frac{N\Pr\{t<D\}}{T},
	\end{equation}
	where $N/T$ denotes the number of generated tasks in unit time.	The traffic efficiency is the traffic flow rate, i.e., the average number of vehicles passing through in unit time, which is 
	\begin{equation}
	    \eta_\textrm{TE} = VL.
	\end{equation}
	To study the tradeoff between $\eta_\textrm{CE}$ and $\eta_\textrm{TE}$, the Fig. \ref{Fig_ct} is depicted. In the win-win region, the eVCC and traffic efficiency both go up with the increased vehicle density. It is observed that after some point, which actually corresponds to $L^\dag$ in Corollary \ref{prop1}, the eVCC efficiency drops with the traffic efficiency, which is denoted by the tradeoff region in Fig. \ref{Fig_ct}. The finding indicates that the future intelligent transportation system should strike a good balance between traffic efficiency and eVCC efficiency.
	
	\begin{figure*}[!ht]
    	\centering
    	\includegraphics[width=0.7\textwidth]{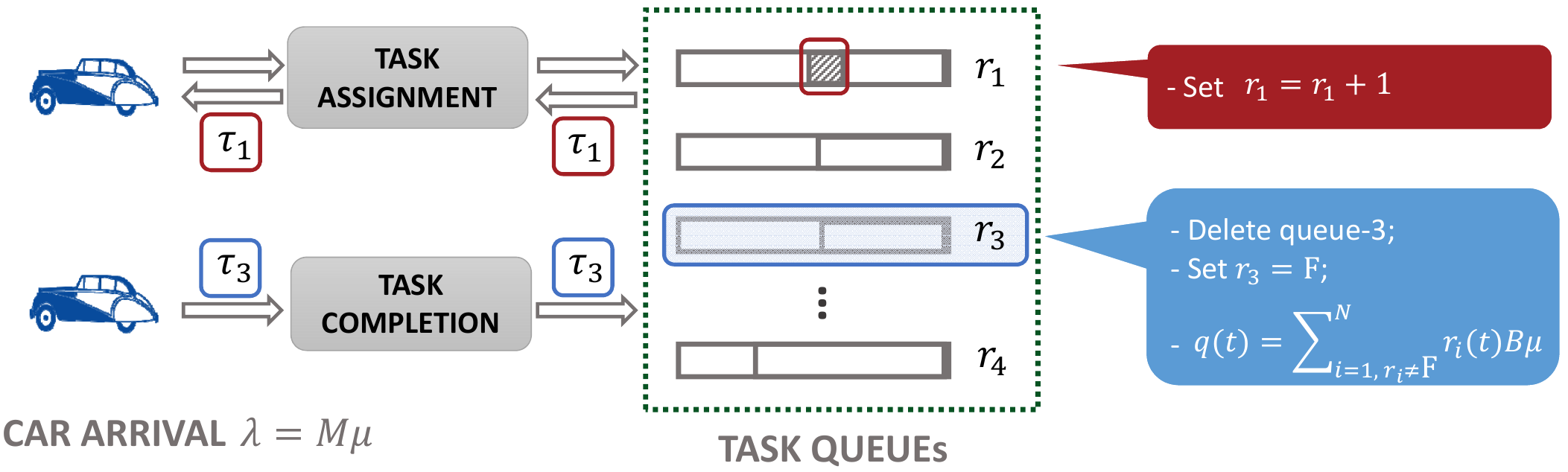}
    	\caption{Pictorial description of MDP-2.The discrete meeting probability is denoted by $p$, and the departure probability is denoted by $q$ which are both i.i.d. over time and follows Bernoulli distribution.}
    	\label{Fig_mdp2}
    \end{figure*}
	\section{Conclusions}
	\label{sec_cl}
	In this paper, in order to minimize the deadline violation probability in the eVCC system, we design optimal multi-task replication policy by finite horizon MDP. Further, it is proved that the optimal policy given homogeneous memory-less vehicle meeting and identical deadline of tasks is the BETA policy. The BETA policy always assigns the task with the least number of replicas, hence the longest expected time to go, which happens to be the opposite of several existing deadline-driven algorithms \cite{sun15} \cite{guo15} \cite{schr68} for applications with different contexts. 
	
	A tight, closed-form upper bound for the deadline violation probability by the BETA policy is derived which shows that the execution time is Rayleigh distributed with variance scaling with $\lambda V^2$. It is intriguing to find out that, combining with the traffic engineering theory, the optimum vehicle density to minimize the deadline violation probability is lower than the traditional critical vehicle density which maximizes the traffic flow rate. Given that the average vehicle speed decreases as the vehicle density grows, the result implies that the increased vehicle mobility benefits the eVCC system more compared to the traditional traffic system attributing to higher vehicle meeting rate, until the point that there are too few vehicles on the road due to decreased vehicle density. Such a tradeoff is thoroughly described by the presented analysis. A general and sufficient condition under which the optimum vehicle speed for the eVCC system is larger than that for the traffic system is given. Combining with the linear speed-density relationship and the asymptotic analysis, it is shown that the optimum vehicle density for eVCC is one-third of the traffic jam vehicle density. Simulation results show that the derived closed-form expressions can fully characterize the impacts of several key parameters.
	
	The future directions include extensions to non-exponential service time distributions and non-identical task deadlines. With task service time no longer being memory-less, the task with a longer waiting time currently should expect the waiting time to be shorter and even longer, with light-tail and heavy-tail service time distributions, respectively. The task replication policy is hence affected. When the tasks have different deadlines, it is natural that they should be treated differently, which deviates from the BETA policy. In either case, it is conjectured that the BETA policy can be extended by leveraging a utility function \cite{hou12} to characterize those factors.  
	\appendices
	\section{Proof of Theorem \ref{thm1}}
	\label{app_thm1}
	\begin{IEEEproof}
		The sketch of the proof is as follows. First, we will derive an equivalent MDP which has the same optimal policy as the above-mentioned MDP. Then we will show that the BETA policy is optimal based on the equivalent MDP. Towards developing the equivalent MDP, we first prove the following lemma.
		
		\begin{lemma}
			\label{lm_eq}
			Define the total service rate under policy $\bm{\pi}$ as 
			\begin{equation}
			R(\bm{\pi})=\sum_{t=0}^{D-1}\sum_{i=1, r_{i,\bm{\pi}}(t)\neq \textrm{F}} ^{N} r_{i,\bm{\pi}}(t)\mu B.
			\end{equation}
			If $R(\bm{\pi}_1)\geq R(\bm{\pi}_2)$, the distribution of $p_\textrm{v}(\bm{\pi}_1)$ dominates that of $p_\textrm{v}(\bm{\pi}_2)$, i.e., 
			\begin{equation}
			    \Pr\{p_\textrm{v}(\bm{\pi}_1)<x\}\geq \Pr\{p_\textrm{v}(\bm{\pi}_2)<x\}.
			\end{equation} 
			Consequently, the average violation ratio under $\bm{\pi}_1$ is no more than that under $\bm{\pi}_2$.
		\end{lemma}
		
		\begin{IEEEproof}
    		Lemma \ref{lm_eq} simply implies that if a policy has a stronger serving ability (higher service rate), it has better violation ratio. To make it rigorous, a coupling argument is adopted. Consider a given total service rate of $R(\bm{\pi})$. The service process is with a time varying service rate of $r_i(t)B$, $i\in \{1,\cdots, N\}$. $r_i(t)\neq \textrm{F}$, and the serving time is from $0$ to $D$. Given that the service process is memory-less and independent with the policy once the service rate is given. The considered service process is equivalent with a system with $R(\bm{\pi})$ servers, and service rates are all $1$, serving for one time slot. The equivalence stem from the fact that the distribution of the number of service tasks are the same.
    		
    		Due to the i.i.d. assumption, we can consider that the servers under $\bm{\pi}_2$ are a subset of the servers under $\bm{\pi}_1$, given that $R(\bm{\pi}_1)\geq R(\bm{\pi}_2)$. Based on a coupling argument, assuming the occurrence of an arbitrary, and identical subset of servers with task completion in the time slot, the number of task completion under $\bm{\pi}_1$ is higher than that under $\bm{\pi}_2$, since the servers under $\bm{\pi}_2$ is a subset of $\bm{\pi}_1$. This coupling argument results in the distribution dominance. The average results follow immediately. With this, we complete the proof. 
    	\end{IEEEproof}
		
		Next, we develop an equivalent MDP (denoted by MDP-2), based on the above-mentioned MDP (denoted by MDP-1).
		
		\begin{lemma}
			\label{lm_e}
			The optimal policy of MDP-1 is the same with the following MDP-2.
		\end{lemma}
		
		The system is still time-slotted as previously explained in Section \ref{sec_sm}. After the task assignment, the vehicle is placed in one of the $N$ queues (see Fig. \ref{Fig_mdp2} for an illustration), where each queue denotes a task with $r_i$ task replicas. The $N$ queues are served by a common server with a service rate of $\sum_{i=1,r_i(t)\neq \textrm{F}}^{N}r_i(t)B\mu$. Whenever a task completion occurs at time $t_\textrm{m}$, the corresponding queue is deleted, and we set $r_i(t)=\textrm{F}, \forall t>t_\textrm{m}$. The total reward is defined as 		
		\begin{equation}
		\mathbb{E}\left[\sum_{t=0}^{D-1} \sum_{i=1,r_i(t)\neq \textrm{F}}^N r_i(t)\right].
		\end{equation}
		\begin{IEEEproof}
			By construction of MDP-$2$ and Lemma \ref{lm_e}, it is straightforward to see that the optimal policies of MDP-$1$ and MDP-$2$ are the same. 
		\end{IEEEproof}
		
		Based on Lemma \ref{lm_e}, in order to show that the BETA policy is optimal for MDP-1, it is sufficient to show that it is optimal for MDP-2. Towards this end, we will first show that the BETA policy is the myopic policy for MDP-2, meaning that the BETA policy maximizes the immediate reward for MDP-2. Then, we will show that the myopic policy is indeed the optimal policy, and hence the BETA policy is optimal for MDP-1. Similar techniques have been used in \cite{zhao08}, but in a different scenario.
		
		Define $g_{t}(\bm{r}(t),a)$ as the immediate reward, given the queue status $r(t)$, and the action of assigning to the $a$-th queue. Denote by $c(t)=1$ as that there is a vehicle meeting with the task-RSU, and $c(t)=0$ otherwise. Then, based on the law of total probability,	
		
		\begin{IEEEeqnarray}{rCl}
			\label{eqn:EEE}
			&& \mathbb{E}\left[g_{t}(\bm{r}(t),a)\right] \nonumber\\
			&=& \left(1-
			\sum^N_{i=1,r_i(t)\neq \textrm{F}}
			p_i\right) \left(
			\sum^N_{i=1,r_i(t)\neq \textrm{F}}
			r_i(t)+c(t)\right) \nonumber\\
			&& + \sum^N_{i=1,r_i(t)\neq \textrm{F}}
			p_i \left(
			\sum^N_{n=1,r_n(t) \neq \textrm{F}}
			r_n(t)+c(t) \right. \nonumber\\
			&&  \qquad \qquad \qquad \qquad \qquad \qquad \qquad  \left.-r_i(t)-\mathds{1}(a=i)c(t) \right)  \nonumber\\
			&=&\sum^N_{i=1,r_i(t)\neq \textrm{F}}
			r_i(t)+c(t) \nonumber\\
			&& \qquad \qquad \qquad \,\,\,\, - \sum^N_{i=1,r_i(t)\neq \textrm{F}}
			p_i (r_i(t)+\mathds{1}(a=i)c(t)), \nonumber\\
		\end{IEEEeqnarray}		
		where $p_i$ denotes the probability that a task which belongs to queue-$n$ is executed in this time slot.		
		\begin{equation}
		\label{eqn:p_iii}
		p_i=(r_i(t)+\mathds{1}(i=a)c(t))\mu B\delta, \forall 1\leq i\leq N, r_i(t)\neq \textrm{F}.
		\end{equation}
		
		After a task completion from queue-$n$, the queue-$n$ is deleted. Based on (\ref{eqn:EEE}) and (\ref{eqn:p_iii}), it is easy to see that the myopic policy which maximizes $\mathbb{E}[g_{t}(\bm{r}(t),a)]$ is to assign the task-$n_\textrm{m}$ to the vehicle, where		
		\begin{equation}
		n_\textrm{m}=\argmin_
		{i\in\{1,\cdots,N\}, r_i(t)\neq \textrm{F}} 
		r_i(t).
		\end{equation}
		This coincides with the BETA policy. 
		
		In what follows, we will show the myopic, i.e., BETA policy, is optimal. The proof is based on backwards induction on time $t$. Define the total reward from time $t$ on based on the BETA policy given the current $\bm{r}(t)$ as $W_t(\bm{r}(t))$. The following statements will be shown to be valid based on backwards induction.		
		\begin{itemize}
			\item[I)] $W_t(\bm{r}(t))$ is the optimal total reward from $t$ on, i.e., the BETA policy is optimal.
			\item[II)] For each $t\geq 0$, if $r_i(t)\leq r_j(t)$
			\begin{equation}
			    W_t(\bm{r}(t)\textrm{ with }r_i(t)=\textrm{F}) \geq W_t(\bm{r}(t)\textrm{ with }r_j(t)=\textrm{F}).
			\end{equation}
		\end{itemize}
		
		For notation simplicity, denote $W_t(\bm{r}(t)$ with $r_i(t)=\textrm{F})$ as $\hat{W}_{t,i}(\bm{r}(t))$. Given that I) and II) are both valid from time $t+1$ to $D$. We want to show that they are also valid at time $t$. Define the total reward from time $t$ on, given the action $a$, and followed by actions based on the BETA policy from $t+1$ on, as $q_{t}(\bm{r}(t),a)$. Then I) is equivalent to 		
		\begin{equation}
		W_t(\bm{r}(t)) = \max_a q_{t}(\bm{r}(t),a),
		\end{equation}
		where $q_{t}(\bm{r}(t),a)$ is given in \eqref{qq} at the top of next page.
		\begin{figure*}
			\begin{IEEEeqnarray}{rCl}
			\label{qq}
				q_{t}(\bm{r}(t),a) &=& \mathbb{E}\left[g_{t}(\bm{r}(t),a)\right] + \sum^N_{i=1,r_i(t)\neq \textrm{F}}
				p_i W_{t+1}(\bm{r}(t)) + \sum^N_{i=1,r_i(t)\neq \textrm{F}}
				p_i \hat{W}_{t+1,i}(\bm{r}(t)) \nonumber\\
				&=&\sum^N_{i=1,r_i(t)\neq \textrm{F}}
				r_i(t)+c(t) - \sum^N_{i=1,r_i(t)\neq \textrm{F}}
				(r_i(t)+\mathds{1}(a=i)c(t))^2\mu B\delta+ \sum^N_{i=1,r_i(t)\neq \textrm{F}}
				p_i W_{t+1}(\bm{r}(t))   \nonumber \\
				&& + \sum^N_{i=1,r_i(t)\neq \textrm{F}} (r_i(t)+\mathds{1}(a=i)c(t))\mu B\delta \hat{W}_{t+1,i}(\bm{r}(t)).   
			\end{IEEEeqnarray}
		\end{figure*}
		It follows that		
		\begin{IEEEeqnarray}{rCl}
			a_\textrm{m} &=& \argmax_a q_{t}(\bm{r},a) \nonumber\\
			&=& \argmax_a (-2r_a(t)+\hat{W}_{t+1,a}(\bm{r}(t))).
		\end{IEEEeqnarray}		
		Based on the induction hypothesis in II), it follows that 		
		\begin{IEEEeqnarray}{rCl}
			a_\textrm{m} &=& \argmax_a (\hat{W}_{t+1,a}(\bm{r},a)) \nonumber\\
			&=& \argmin_a [r_a(t)].
		\end{IEEEeqnarray}		
		With this, we establish the fact that the BETA policy is optimal at time $t$, given the induction hypothesis in I) and II).
		
		To show II) is also valid at time $t$, it follows that given $r_i\leq r_j$ and $m=\argmin_{1\leq n\leq N}r_n(t)$, $\forall i,j\in\{1,\cdots,N\}$.			
		\begin{IEEEeqnarray}{rCl}
			&&\hat{W}_{t,i}(\bm{r}(t))-\hat{W}_{t,j}(\bm{r}(t))   \nonumber\\
			\label{d1}
			&=& r_j-r_i+\left((r_j+\mathds{1}(j=m)c(t))^2 - (r_i+\mathds{1}(i=m)c(t))^2 \right)\nonumber \\
			&&\\
			\label{d2}
			&&+ p_j\hat{W}_{t+1,i}(\bm{r}(t))-p_i\hat{W}_{t+1,j}(\bm{r}(t)) \nonumber\\
			&& \qquad \qquad + \sum^N_{i=1,r_i(t)\neq \textrm{F}} \left(\hat{W}_{t+1,i}(\bm{r}(t))-\hat{W}_{t+1,j}(\bm{r}(t))\right)\\
			\label{d3}
			&&+ \sum_{n\neq i} p_n\hat{W}_{t+1,n,i}(\bm{r}(t))- \sum_{n\neq j} p_n\hat{W}_{t+1,n,j}(\bm{r}(t)).
		\end{IEEEeqnarray}	
		
		It is easy to see that \eqref{d1} $\geq 0$, since $r_j\geq r_i$. 
		
		To show \eqref{d2}$\geq 0$, two cases are considered. If $r_i=r_j$, then \eqref{d2} is trivially true. Otherwise given $r_j\geq r_i$, $r_i+1\leq r_j$, then		
		\begin{IEEEeqnarray}{rCl}
			p_j &=& (r_j+\mathds{1}(m_j=j)c(t))\mu B\delta \geq (r_i+1) \mu B\delta \nonumber\\
			&\geq& (r_i + \mathds{1}(m_i=i)c(t))\mu B\delta = p_i,
		\end{IEEEeqnarray}
		where 
		\begin{IEEEeqnarray}{rCl}
		m_j &=& \argmin_{n\neq j}r_n(t), \nonumber\\
		m_i &=& \argmin_{n\neq i} r_n(t). 
		\end{IEEEeqnarray}
		Combining with the induction hypothesis II), we obtain \eqref{d2}$\geq 0$.
		
		To show that \eqref{d3}$\geq 0$, let us consider the following four cases. 
		
		Case A: $m_i\neq j$, $m_j\neq i$. Then, $m_i=m_j$.		
		\begin{IEEEeqnarray}{rCl}
			\eqref{d3}&=&r_j(t)\mu B\delta \hat{W}_{t+1,i,j}(\bm{r}(t)) \nonumber\\
			&&-r_i(t)\mu B\delta\hat{W}_{t+1,i,j}(\bm{r}(t))                 \nonumber\\
			&&+ \sum_{n\neq i,j}p_n\left(\hat{W}_{t+1,n,i}(\bm{r}(t))-\hat{W}_{t+1,n,j}(\bm{r}(t))\right) \nonumber\\
			&&\geq 0,
		\end{IEEEeqnarray}
		based on $r_j(t)\geq r_i(t)$ and induction hypothesis II).
		
		Case B: $m_i=j$, $m_j\neq i$. Then, $r_i(t)\leq r_j(t)=r_{m_i}(t)\leq r_n$, $\forall n\neq i,j$. It follows that $m_j=i$, and this case is infeasible.
		
		Case C: $m_i\neq j, m_j =i$. 		
		\begin{IEEEeqnarray}{rCl}
			\eqref{d3} &=&(r_{m_i}(t)+c(t))\mu B\delta\hat{W}_{t+1,m_i,j}(\bm{r}(t))  \nonumber\\
			&&+ r_j(t)\mu B\delta \hat{W}_{t+1,i,j}(\bm{r}(t))                            \nonumber\\
			&&- (r_i(t)+c(t))\mu B\delta\hat{W}_{t+1,i,j}(\bm{r}(t))                     \nonumber\\
			&&- r_{m_i}(t)\mu B\delta\hat{W}_{t+1,m_i,j}(\bm{r}(t)) \geq 0,
		\end{IEEEeqnarray}
		based on the same technique to prove \eqref{d2}$\geq 0$.
		
		Case D: $m_i=j$, $m_j=i$.
		\begin{IEEEeqnarray}{rCl}
			\eqref{d3}&=&(r_j(t)+c(t))\mu B\delta\hat{W}_{t+1,i,j}(\bm{r}(t)) \nonumber\\
			&&- (r_i(t)+c(t))\mu B\delta\hat{W}_{t+1,i,j}(\bm{r}(t))\geq 0. 
		\end{IEEEeqnarray}
		
		With this, we complete the induction proof, and hence the theorem is proved.
	\end{IEEEproof}
	
	\section{Proof of Theorem \ref{thm2}}
	\label{app_thm2}
	\begin{IEEEproof} 
	We prove the theorem by first showing that the deadline violation probability is upper bounded by a system with one task and $\lambda S/N$ vehicle meeting rate. Then the performance of the task system is derived in a closed-form.
		
		\begin{lemma} 
			\label{lm1}
			The deadline violation probability is upper bound by a system with one task and a vehicle meeting rate of $\lambda S\mu/N$.
		\end{lemma}
		\begin{IEEEproof} 
		Consider a round-robin task assignment policy which assigns the $(k\,\textrm{mod}\,N)$-th task to the $k$-th meeting vehicle, even if the task has been finished. Since the division of a Poisson process is also a Poisson process, the resultant meeting rate in MDP-2 for the $k$-th queue is $\lambda S/N$. It is straightforward to see the performance of the BETA policy is upper bounded by this round-robin policy by noticing that the BETA policy is the same with the round-robin policy, except that it does not assign finished tasks.
		\end{IEEEproof}
		\begin{lemma}
			\label{lm2}
			For a system with one task and a vehicle meeting rate of $\lambda S\mu/N$, the deadline violation probability is identical to a system with $\lambda S/N$ servers each of which has an i.i.d. service time distribution of 		
			\begin{equation}
			\label{eqn:Fx}
			F(x) = 1+\frac{\lambda_1}{\lambda_2-\lambda_1}e^{-\lambda_2 x}-\frac{\lambda_2}{\lambda_2-\lambda_1}e^{-\lambda_1 x}.
			\end{equation}	
			The task is completed once one of the servers finishes.
		\end{lemma}
		\begin{IEEEproof} 
		Since there is only one task, all the meeting vehicles are carrying the same task. The total number of vehicles is $\lambda S/N$ (ignoring the remainder of division), and each vehicle needs an extra service time of $t_\textrm{s}$ after meeting at the task-RSU, where 
		\begin{equation}
		t_\textrm{s}\sim \textrm{Exp}(B\mu).
		\end{equation}
		Therefore, the service process is equivalent to one with $\lambda S/N$ servers, and each with a service time.			
		\begin{equation}
		t=t_\textrm{a}+t_\textrm{s},
		\end{equation}
		where 
		\begin{equation}
		    t_\textrm{a}\sim \textrm{Exp}(\mu)
		\end{equation}
		denotes the meeting delay distribution, $t_\textrm{a}$ and $t_\textrm{s}$ are independent.
		
		The natural way to establish the distribution of the sum of two independent random variables is to leverage the characteristic function (CF). Given the CF of a random variable $x \sim \textrm{Exp}(\lambda)$ as 		
		\begin{equation}
		\varphi_x(z) = \lambda(\lambda-jz)^{-1},
		\end{equation}
		where $j=\sqrt{-1}$ here, and the CF of $t$ is hence 
		\begin{IEEEeqnarray}{rCl}
			\varphi_t(z)&=&\varphi_{t_\textrm{a}}(z)\cdot\varphi_{t_\textrm{s}}(z)    \nonumber\\
			&=&\lambda_1\left(\lambda_1-jz)^{-1}\lambda_2(\lambda_2-jz\right)^{-1} \nonumber\\
			&=& \left(\frac{1}{\lambda_1-jz}-\frac{1}{\lambda_2-jz}\right)\frac{\lambda_1\lambda_2}{\lambda_2-\lambda_1} \nonumber\\
			&=& \lambda_1(\lambda_1-jz)^{-1}\cdot \frac{\lambda_2}{\lambda_2-\lambda_1} \nonumber\\
			&& + \lambda_2 (\lambda_2-jz)^{-1}\cdot\frac{\lambda_1}{\lambda_1-\lambda_2}.
		\end{IEEEeqnarray}
		It follows that the density function of $t$ is		
		\begin{equation}
		f(t) = \frac{\lambda_2}{\lambda_2-\lambda_1} \lambda_1 e^{-\lambda_1 t} + \frac{\lambda_1}{\lambda_1-\lambda_2} \lambda_2 e^{-\lambda_2 t},
		\end{equation}	
		and hence the distribution function follows immediately.
		\end{IEEEproof}
		
		Based on Lemma \ref{lm1} and Lemma \ref{lm2}, the violation happens when all servers fail to execute the task on time, i.e.,		
		\begin{equation}
		t_\textrm{m} \triangleq \min_{i\in \{1,\cdots,M\}}[t_i]>D,
		\end{equation}
		where $M$ is a Poisson distributed random variable with mean of $\lambda S/N$ to follow the assumption 1) in Section \ref{sec_sm}. The deadline violation probability follows, which reads 		
		\begin{IEEEeqnarray}{rCl}
			\Pr\{t_\textrm{m}>D\} &=& \sum_{m=0}^{\infty}\Pr\{t_\textrm{m}>D|M=m\}\Pr\{M=m\}. \nonumber\\
		\end{IEEEeqnarray}
		To proceed, we have the following lemma.
		
		\begin{lemma} 
			\label{lm3}
			Given a distribution function $F(x)$,
			\begin{equation}
			v(x) \triangleq \sum_{k=0}^{\infty}\frac{1}{k!}\alpha^k e^{-\alpha}(1-F(x))^k = e^{-\alpha F(x)}.
			\end{equation}
		\end{lemma}
		
		\begin{IEEEproof} 			
			\begin{IEEEeqnarray}{rCl}
				v(x)&=&e^{-\alpha}+\sum_{k=1}^{\infty}\frac{1}{k!}e^{-\alpha}\alpha^k(1-F(x))^k,\\
				\frac{\textrm{d}v(x)}{\textrm{d}x}&=& -\sum_{k=1}^{\infty}\frac{1}{k!}e^{-\alpha} \alpha^k (1-F(x))^{k-1} \frac{\textrm{d}F(x)}{\textrm{d}x} \nonumber\\
				&=& -\alpha \sum_{k=0}^{\infty} \frac{1}{k!}e^{-\alpha}\alpha^k (1-F(x))^k \frac{\textrm{d}F(x)}{\textrm{d}x}    \nonumber\\
				&=&-\alpha \frac{\textrm{d}F(x)}{\textrm{d}x} v(x).
			\end{IEEEeqnarray}
			
			Therefore,			
			\begin{IEEEeqnarray}{rCl}
				\frac{\textrm{d}v(x)}{v(x)} &=& -\alpha \textrm{d}F(x), \\
				v(x) &=& C e^{-\alpha F(x)}.
			\end{IEEEeqnarray}
			
			Since when $x=0$, $F(0)=0$, then $v(0)=1$, $C=1$, and $v(x)=e^{-\alpha F(x)}$.
		\end{IEEEproof}
		
		Based on Lemma \ref{lm3}, 		
		\begin{equation}
		\Pr\{t_\textrm{m}>D\} = e^{-\frac{\lambda S}{N}F(x)},
		\end{equation}
		where $F(x)$ is given in (\ref{eqn:Fx}), which concludes the proof.
	\end{IEEEproof}
	\section{Proof of Corollary \ref{coro_ray}}
	\label{app_coro_ray}
	\begin{IEEEproof}
	Consider the Taylor series
	\begin{equation}
	e^{-\lambda x} = 1-\lambda x + \frac{1}{2}\lambda^2 x^2 + o(x^2),
	\end{equation}
	where $o(x^n)$ denotes higher order infinitesimals of $x^n$. When $\lambda_1 D$ and $\lambda_2 D$ approach zeros, the upper bound in Theorem \ref{thm2} is asymptotically
	\begin{IEEEeqnarray}{rCl}
		&& \exp \left\{-\frac{\lambda S}{N}\left(1+\frac{\lambda_1}{\lambda_2-\lambda_1}e^{-\lambda_2 D} -\frac{\lambda_2}{\lambda_2-\lambda_1}e^{-\lambda_1 D}\right)\right\} \nonumber \\
		&\xrightarrow{D \to 0}& \exp \left\{-\frac{\lambda S}{N}\left(1+\frac{\lambda_1}{\lambda_2-\lambda_1} \left(1-\lambda_2 x + \frac{1}{2}\lambda_2^2 D^2 \right)\right.\right. \nonumber \\
		&& \left. \left. -\frac{\lambda_2}{\lambda_2-\lambda_1}\left(1-\lambda_1 D + \frac{1}{2}\lambda_1^2 D^2\right)  + o(D^2) \right)\right\} \nonumber\\
		& = & \exp\left\{-\frac{\lambda S}{2N} \lambda_1 \lambda_2 D^2 + o(D^2)\right\}.
	\end{IEEEeqnarray}
	The conclusion follows immediately by noticing $\lambda_1 = \mu$ and $\lambda_2 = B \mu$, which completes the proof.
	\end{IEEEproof}
	\section{Proof of Theorem \ref{thm3}}
	\label{app_thm3}
	\begin{IEEEproof}
	First, it is easy to see that 
	\begin{IEEEeqnarray}{rCl}
	    L^* &=& \argmax_{0 \le L \le L_\textrm{max}}[V(L)L],\\
	    L^\dag &=& \argmax_{0 \le L \le L_\textrm{max}}[V^2(L)L].
	\end{IEEEeqnarray}
	It is obvious that both $L^*$ and $L^\dag$ are positive. If $L^*=L_\textrm{max}$, then $L^\dag \le L^*$ is trivially true. Otherwise, we have, based on the first-order Karush-Kuhn-Tucker (KKT) condition,
	\begin{IEEEeqnarray}{rCl}
	V(L^*)+L^* V^\prime(L^*) &=& 0,\\
	V(L^\dag)+2L^\dag V^\prime(L^\dag) &=& 0.
	\end{IEEEeqnarray}
	It is easy to find out that if $V(L)$ is concave, then $V^\prime(L)$ is non-increasing, and therefore $L^* \ge L^\dag$. Otherwise if $V(L)$ satisfies the condition in \eqref{v_condition2}, wherein $V(L)$ has a polynomial form, it is sufficient to show for each monomial term, $L^* \ge L^\dag$. Specifically, we consider
	\begin{equation}
	    V(L) = g -  c L^{\alpha},
	\end{equation}
	where $\alpha \in (0,+\infty)\cup(-1,-\frac{1}{2})$, $c \ge 0$, $\forall k$ and $g \ge 0$. It follows based on the KKT condition that 
	\begin{IEEEeqnarray}{rCl}
	L^* &=& \left(\frac{g}{c(1+\alpha)}\right)^{\frac{1}{\alpha}},\\
	L^\dag &=& \left(\frac{g}{c(1+2\alpha)}\right)^{\frac{1}{\alpha}}.
	\end{IEEEeqnarray}
	Hence $L^* \ge L^\dag$ is equivalent to 
	\begin{equation}
	\frac{g\alpha}{c(1+2\alpha)(1+\alpha)} \ge 0.
	\end{equation}
	The condition in \eqref{v_condition2} follows immediately, which concludes the proof.
	\end{IEEEproof}
	\section{Proof of Corollary \ref{prop1}}
	\label{app_coro2}
	In our model, the vehicle density is $\lambda$. The rate of traffic flow is the reciprocal of the inter-vehicle-meeting time at the task-RSU, i.e., 
	\begin{IEEEeqnarray}{rCl}
		F &=& \mathbb{E}\left[\frac{1}{\min{[t_1,...,t_M]}}\right] = \lambda S \mu,
	\end{IEEEeqnarray}
	where $t_i$ denotes the inter-meeting time of the $i$-th vehicle. Therefore, the average speed is 
	\begin{IEEEeqnarray}{rCl}
		V &=& \frac{F}{\lambda} = S \mu.
	\end{IEEEeqnarray}
	Combining with \eqref{VD}, we obtain
	\begin{IEEEeqnarray}{rCl}
		\label{mu_l}
		\mu = \frac{V_\textrm{max}}{S}\left(1-\frac{\lambda}{L_\textrm{max}}\right).
	\end{IEEEeqnarray}
	Given that in the short deadline regime, the deadline violation probability follows the Rayleigh distribution given in Corollary \ref{coro_ray}, the optimum vehicle density is therefore 
	\begin{equation}
	L^\dag = \argmax_{0 \le \lambda \le L_\textrm{max}}[\lambda \mu^2].
	\end{equation}
	Combining with \eqref{mu_l}, the conclusion of the proposition follows immediately based on the Cauchy-Schwarz inequality.
	
	\bibliographystyle{ieeetr}
	\bibliography{eVCC}
\end{document}